\documentstyle[emulapj]{article}
\begin{document}

\lefthead{HYDRODYNAMIC SIMULATIONS OF COUNTERROTATING DISKS}
\righthead{KUZNETSOV ET AL.}

\title{HYDRODYNAMIC SIMULATIONS OF COUNTERROTATING ACCRETION
DISKS}

\author{O.A.Kuznetsov\altaffilmark{1}}
\affil{Keldysh Institute of Applied Mathematics, Russian Academy
of Sciences, Moscow, Russia}

\author{R.V.E.Lovelace\altaffilmark{2}}
\affil{Department of Astronomy, Cornell University, Ithaca, NY
14853-6801}

\author{M.M.Romanova\altaffilmark{3}}
\affil{Space Research Institute, Russian Academy of Sciences,
Moscow, Russia; and\\ Department of Astronomy, Cornell
University, Ithaca, NY 14853-6801}

\and

\author{V.M.Chechetkin\altaffilmark{4}}
\affil{Keldysh Institute of Applied Mathematics, Russian Academy
of Sciences, Moscow, Russia}

\altaffiltext{1}{E-mail: {\it kuznecov@spp.keldysh.ru}}
\altaffiltext{2}{E-mail: {\it rvl1@cornell.edu}}
\altaffiltext{3}{E-mail: {\it romanova@astrosun.tn.cornell.edu}}
\altaffiltext{4}{E-mail: {\it chech@int.keldysh.ru}}

\bigskip

\submitted{}

\slugcomment{Accepted for publication in the Astrophysical Journal}

\begin{abstract}

Time-dependent, axisymmetric hydrodynamic simulations have been
used to study accretion disks consisting of counterrotating
components with an intervening shear layer(s).  Configurations
of this type can arise from the accretion of newly supplied
counterrotating matter onto an existing corotating disk.  The
grid-dependent numerical viscosity of our hydro code is used to
simulate the influence of a turbulent viscosity of the disk.
Firstly, we consider the case where the gas well above the disk
midplane ($z>0$) rotates with angular rate $+\Omega(r)$ and that
well below ($z<0$) has the same properties but rotates with rate
$-\Omega(r)$.  We find that there is angular momentum
annihilation in a narrow equatorial boundary layer in which
matter accretes supersonically with a velocity which approaches
the free-fall velocity.  This is in accord with the analytic
model of Lovelace and Chou (1996).  The average accretion speed
of the disk can be enormously larger than that for a
conventional $\alpha-$disk rotating in one direction.  Under
some conditions the interface between the corotating and
counterrotating components shows significant warping.  Secondly,
we consider the case of a corotating accretion disk for $r <
r_t$ and a counterrotating disk for $r> r_t$.  In this case we
observed, that matter from the annihilation layer lost its
stability and propagated inward pushing matter of inner regions
of the disk to accrete.  Thirdly, we investigated the case where
counterrotating matter inflowing from large radial distances
encounters an existing corotating disk.  Friction between the
inflowing matter and the existing disk is found to lead to fast
boundary layer accretion along the disk surfaces and to enhanced
accretion in the main disk.

These models are pertinent to the formation of counterrotating
disks in galaxies and possibly in Active Galactic Nuclei and in
X-ray pulsars in binary systems.  For galaxies the high
accretion speed allows counterrotating gas to be transported
into the central regions of a galaxy in a time much less than
the Hubble time.
\end{abstract}


\section{Introduction}

The widely considered models of accretion disks have gas
rotating in one direction with a turbulent viscosity acting to
transport angular momentum outward (Shakura 1973; Shakura \&
Sunyaev 1973).  However, recent observations indicate that there
may be more complicated disk structures in some cases on a
galactic scale, in galactic nuclei, and in disks around compact
stellar mass objects.  Studies of normal galaxies have revealed
counterrotating gas and/or stars in many galaxies of all
morphological types - ellipticals, spirals, and irregulars
(Rubin 1994a, 1994b; Galletta 1995).  In elliptical galaxies,
the counterrotating component is usually in the nuclear core and
may result from merging of galaxies with opposite angular
momentum.

\begin{figure*}[t]
\epsscale{0.4}
\plotone{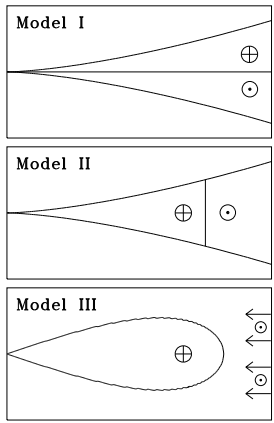}
\caption{
Sketches of three main cases considered in this paper.  The
signs $\oplus$ indicates corotating matter and $\odot$
counterrotating matter.  Details of the models are described in
the text.
}
\end{figure*}

In a number of spirals and S0 galaxies, counterrotating disks of
stars and/or gas have been found to co-exist with the primary
disk out to large distances ($10 - 20$ kpc), with the first
example, NGC 4550, discovered by Rubin, Graham, \& Kenney
(1992).  Of interest here is the ``Evil Eye'' galaxy NGC 4826 in
which the direction of rotation of the gas reverses going from
the inner ($180$ km/s) to the outer disk ($-200$ km/s) with an
inward radial accretion speed of more than $100$ km/s in the
transition zone, while the stars at all radii rotate in the same
direction as the gas in the inner disk which has a radius $\sim
1200$ pc (Braun et al.  1994; Rubin 1994a, 1994b).  The large
scale counterrotating disks probably do not result from mergers
of flat galaxies with opposite spins because of the large
vertical thickening observed in simulation studies of such
mergers (Hernquist \& Barnes 1991; Barnes 1992).  Thakar \&
Ryden (1996) discuss different possibilities, (a) that the
counterrotating matter comes from the merger of an oppositely
rotating gas rich dwarf galaxy with an existing spiral, and (b)
that the accretion of gas occurs over the lifetime of the galaxy
with the more recently accreted gas counterrotating.  Subsequent
star formation in the counterrotating gas may then lead to
counterrotating stars.  Counterrotating gas may have a strong
interaction with the gas resulting from mass loss by corotating
stars (Braun et al. 1994).  Further, the two-stream instability
between counterrotating gas and corotating stars can be
important for gas accretion (Lovelace, Jore, \& Haynes 1996).

Accretion of counterrotating matter onto an existing corotating
disk around a massive black hole may occur in an Active Galactic
Nucleus (AGN).  The counterrotating matter may come from
recently captured material from a galaxy merger, from a tidally
disrupted molecular cloud, or from a star.  Tidal disruption of
a star by the central black hole may release essential part of
the star in the form of gas (Rees 1990; Shlosman, Begelman, \&
Frank 1990; Khokhlov \& Melia 1996).  There is a significant
probability that this gas has angular momentum opposite to that
of the main disk.  This can lead to formation of small disks
with opposite angular momentum in the center of a galaxy (Melia
1994).  This was studied using 3D numerical simulations by
Ruffert (1994), Ruffert \& Melia (1994).  The retrograde disks
may also slow down the rotation of the central black hole
(Moderski \& Sikora 1996).  The greatly enhanced accretion rate
for counterrotating matter in the presence of a corotating disk
may give outbursts of AGN.

Another situation where counterrotating matter may encounter an
existing corotating disk is in low mass X-ray binary sources
where the accreting, magnetized rotating neutron stars are
observed to jump between states where they spin-up and those
where they spin-down.  Nelson et al. (1997a, b) and Chakrabarty
et al. (1997) have proposed that the change from spin-up to
spin-down results from a reversal of the angular momentum of the
wind supplied accreting matter.  Their suggestion is based on
the recent {\it BATSE} observations which show a rapid change
from spin-up to spin-down evolution of some X-ray pulsars.  This
is difficult to explain with conventional theory (see, however,
Lovelace, Romanova, \& Bisnovatyi-Kogan 1998).  Reversal of the
angular momentum of accreted matter in  wind-fed binaries was
observed in the two-dimensional simulations by Blondin et al.
(1990), Livio et al. (1991), and Bisikalo et al. (1996).

An important open problem is to understand the $\it
hydrodynamic$ interaction of counterrotating gaseous disks.
Earlier two-dimensional simulations used either collisionless
particles to represent a system of stars (Davies \& Hunter 1997)
or  ``sticky'' particles (Thakar \& Ryden 1996, and references
therein) to represent a system of clouds.  The sticky particle
interaction gives only a rough representation of a viscous
fluid.  When both components of a counterrotating system are
gaseous, it is important to understand their interaction using
the full equations of hydrodynamics including the viscous
transport of angular momentum.  Hydrodynamic interaction of a
galactic disk with infalling matter of a non-rotating corona was
investigated by Falcke \& Melia (1997) using a one dimensional
approach.  They  showed that this interaction may lead to
significant evolution of the disk.

Lovelace \& Chou (1996) (hereafter LC) developed an analytic
theory of the viscous hydrodynamic interaction of
counterrotating gaseous disks basing on the $\alpha-$viscosity
model of Shakura (1973) and Shakura \& Sunyaev (1973).

Here, we present the results of two-dimensional, time-dependent
hydrodynamic simulations of counterrotating accretion disks.  We
investigate three main models which are sketched in Figure 1.

{\it Model I} is shown in the top sketch of Figure 1: In this
case the accretion disk consists of counterrotating gaseous
components with gas above the disk midplane rotating with
angular rate $+\Omega(r)$, and that below rotating at rate
$-\Omega(r)$. The interface between the components at $z\!\sim\!
0$ constitutes a supersonic shear layer.  A configuration of
this type may arise from the accretion of newly supplied
counterrotating gas onto an existing corotating disk.  For this
case the analytical model of Lovelace \& Chou (1996) proposed
that matter approaching the equatorial plane from above and
below has angular momenta of opposite signs with the result that
there is angular momentum annihilation at $z\sim 0$.  This
matter loses its centrifugal support and accretes at essentially
free-fall speed.

{\it Model II} is shown in the middle sketch in Figure 1:  In
this model we consider that where there is a corotating
($+\Omega$) accretion disk (the ``old'' disk) for $r < r_t$ and
a counterrotating ($-\Omega$) disk (the ``new'' disk) in the
region $r> r_t$.  This case is clearly pertinent to the ``Evil
Eye'' galaxy NGC 4826.

{\it Model III} is shown in the lower sketch of Figure 1: This
model corresponds to inflow of counterrotating matter which
interacts with an existing corotating disk.  In this case
annihilation of angular momentum occurs on the surface of the
existing disk and this can lead to fast accretion.

In \S 2 we discuss the pertinent fluid dynamical equations for
axisymmetric accretion, the initial and boundary conditions, and
the numerical method used.  In \S 3 - \S 5 we discuss results
obtained for Models I-III. We present astrophysical examples in
\S 6.  In \S 7 we give conclusions of this work.

\begin{figure*}[t]
\epsscale{0.5}
\plotone{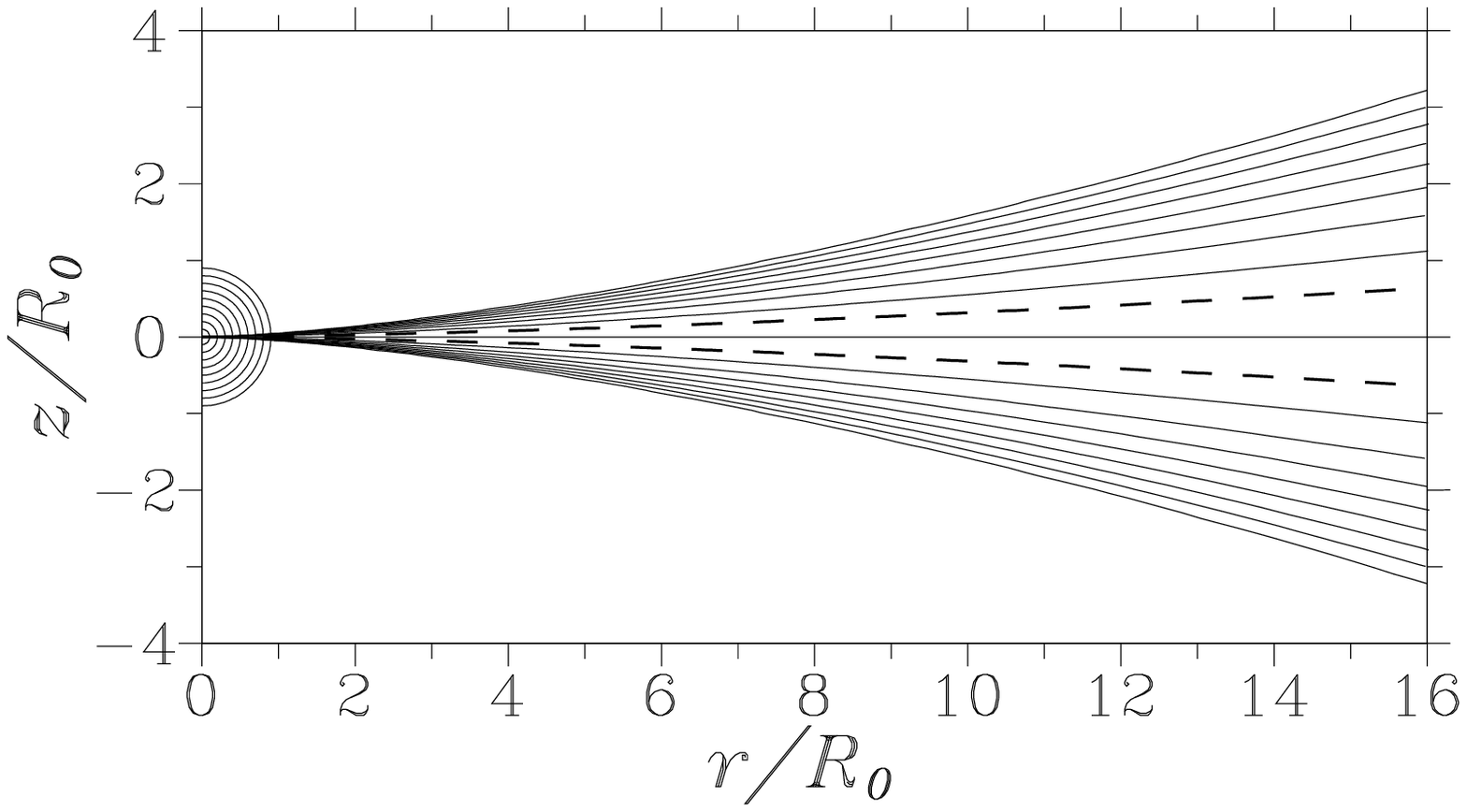}
\caption{
Log-scaled iso-density lines from $\rho_{min}=10^{-6}$ to
$\rho_{max}=1$ for a Keplerian disk described in \S 2.4.1 with
$c_{s0}=0.01$ and $\gamma=1.01$.  The dashed lines show the
standard disk scale-height, $H=rc_s/V_K$.  The circles represent
the region $r^2+z^2 \leq R_0^2$ where matter is considered to be
absorbed.
}
\vspace{0.5cm}
\plotone{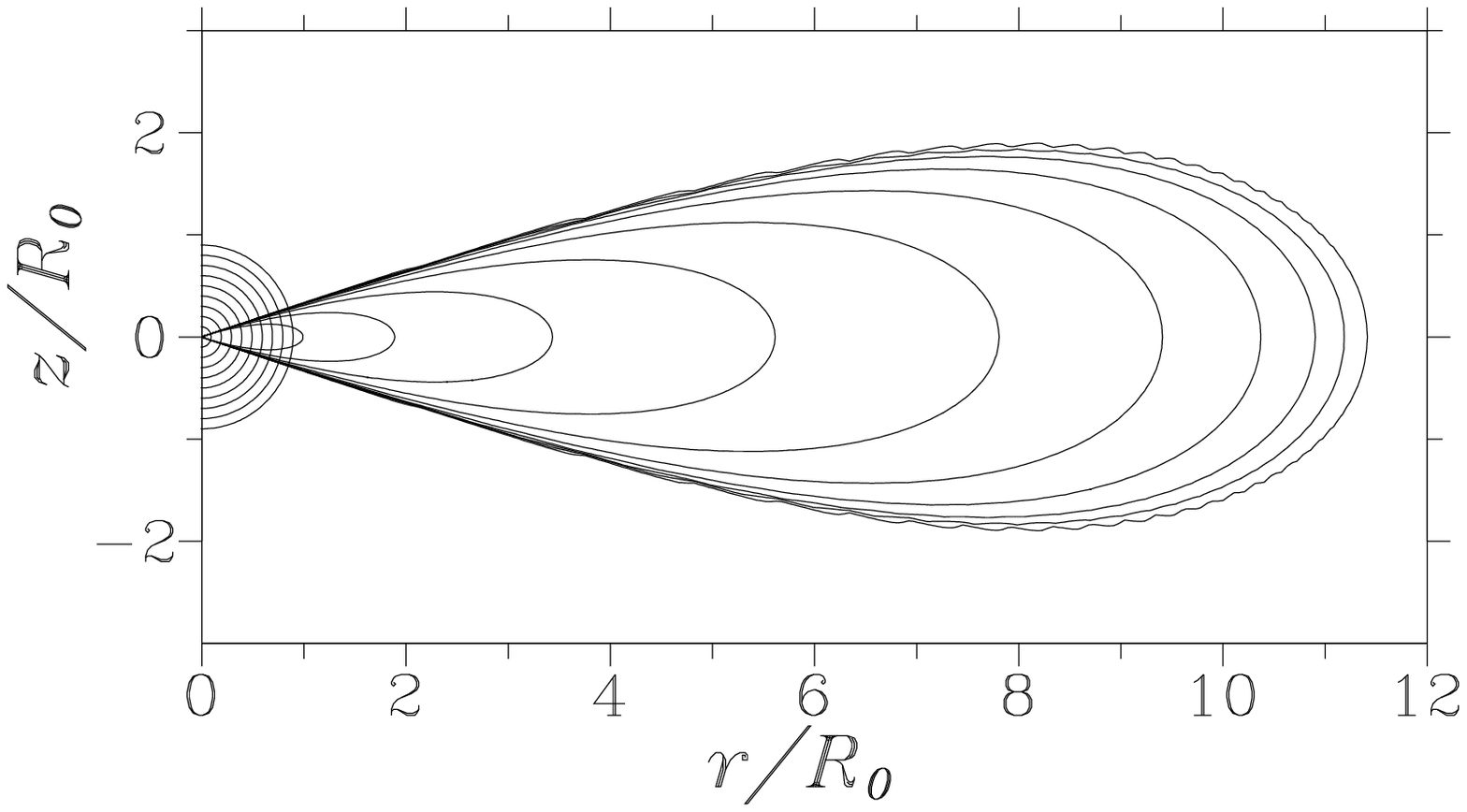}
\caption{
Log-scaled iso-density lines from $\rho_{min}=10^{-6}$ to
$\rho_{max}=1$ for a sub-Keplerian disk as described in \S 2.4.2
for $\Delta = 1/3$, $r_1=11.5R_0$, and $n=3/2$ (or
$\gamma=5/3$).
}
\end{figure*}

\section{Mathematical Model}

\subsection{Basic Equations}

The axisymmetric flows are described by the inviscid
hydrodynamic equations (Euler equations) in spherical, inertial
coordinates $(R,\phi,\theta)$ (with $\theta=0$  the equatorial
plane):

\begin{mathletters}
\begin{equation}
\frac{\partial \rho}{\partial t}
+\frac{1}{R^2}\frac{\partial}{\partial R}
\left(R^2\rho v_R\right)
+\frac{1}{R\cos\theta}
\frac{\partial}{\partial\theta}
\left(\cos\theta\rho v_\theta\right)=0\,,
\end{equation}

$$
\frac{\partial \left(\rho v_R\right)}{\partial t}
+\frac{1}{R^2}\frac{\partial}{\partial R}
\left( R^2(\rho v_R^2+p)\right)~~~~~~~~~~~~~~~~~~~~~~~~
$$
$$
+\frac{1}{R\cos\theta}
\frac{\partial}{\partial\theta}
\left(\cos\theta\rho v_R v_\theta\right)
=
$$
\begin{equation}
=
\frac{2p}{R}-\rho\frac{\partial \Phi_g}{\partial R}
+\rho\frac{v_\phi^2+v_\theta^2}{R}\,,
\end{equation}

$$
\frac{\partial\left(\rho(R\cos\theta v_\phi)\right)}{\partial t}
+\frac{1}{R^2}\frac{\partial}{\partial R}
\left(R^2\rho (R\cos\theta v_\phi) v_R\right)
+
$$
\begin{equation}
+\frac{1}{R\cos\theta}
\frac{\partial}{\partial\theta}
\left(\cos\theta\rho (R\cos\theta v_\phi) v_\theta\right)
=0\,,
\end{equation}

$$
\frac{\partial (\rho v_\theta)}{\partial t}
+\frac{1}{R^2}\frac{\partial}{\partial R}
\left( R^2\rho v_\theta v_R\right)
$$
$$
+\frac{1}{R\cos\theta}
\frac{\partial}{\partial\theta}
\left(\cos\theta(\rho v_\theta^2+p)\right)
=
$$
\begin{equation}
=-\frac{p \tan\theta}{R}-\rho\frac{v_R v_\theta}{R}
-\rho\frac{v_\phi^2}{R}\tan\theta\,,
\end{equation}

$$
\frac{\partial \left(\rho {\cal E}\right)}{\partial t}
+\frac{1}{R^2}\frac{\partial}{\partial R}
\left( R^2\rho \tilde{w} v_R\right)
+\frac{1}{R\cos\theta}
\frac{\partial}{\partial\theta}
\left(\cos\theta\rho \tilde{w} v_\theta \right)
=
$$
\begin{equation}
\label{oa1}
=-\rho\frac{\partial \Phi_g}{\partial R}v_R\,.
\end{equation}
\end{mathletters}

\noindent Here, $~\rho(R,\theta)~$ is the density; ${\bf
v}(R,\theta)$ the flow velocity, with components
$(v_R,v_\phi,v_\theta)$; $p(R,\theta)$ the pressure; $\Phi_g$
the gravitational potential; ${\cal E}$ the full specific energy
${\cal E}\equiv\epsilon+(v_R^2+v_\phi^2+v_\theta^2)/2$;
$\epsilon$ the specific internal energy; and $\tilde{w}$ the
full specific enthalpy, $\tilde{w} \equiv
\epsilon+p/\rho+(v_R^2+v_\phi^2+v_\theta^2)/2$.  An equation of
state is necessary, and we assume the ideal gas equation,

$$
p=(\gamma-1)\epsilon\rho\,,
$$
where $\gamma$ is the usual ratio of heat capacities.  The
gravitational potential is considered to be the Newtonian
potential of the central object.
$$
\Phi_g=-\frac{GM}{R}\,,
$$
That is, we assume that the self gravity of the disk is
negligible.

\subsection{Alternative Form of Energy Equation}

In our early disk simulations we found that the energy equation
taken in divergent form (1e) can lead to inaccurate results.
The reason for this is the large value of the azimuthal velocity
$v_\phi$ and the corresponding kinetic energy $v_\phi^2/2$ in
comparison with other velocities/energies.  The relatively small
quantity $\epsilon$ is determined from (1e) inaccurately in this
case.

To correct this problem we now have excluded the $\phi$-velocity
part of the energy from the energy equation by writing

$$
\frac{\partial}{\partial t}
\left(\rho \frac{v_\phi^2}{2}\right)
+\frac{1}{R^2}\frac{\partial}{\partial R}
\left(R^2\rho \frac{v_\phi^2}{2}v_R\right)+
$$
$$
+\frac{1}{R\cos\theta}
\frac{\partial}{\partial\theta}
\left(\cos\theta\rho \frac{v_\phi^2}{2}v_\theta\right)
=
$$
$$
=-\rho\frac{v_\phi^2 v_R}{R}
+\rho\frac{v_\phi^2 v_\theta}{R}\tan\theta\,,
$$
and subtracting this equation from equation (1e) to obtain
another divergent form of energy equation,

$$
\frac{\partial}{\partial t}\left(\rho {\cal E}'\right)
+\frac{1}{R^2}\frac{\partial}{\partial R}
\left(R^2\rho \tilde{w}' v_R\right)+
$$
$$
+\frac{1}{R\cos\theta}
\frac{\partial}{\partial\theta}
\left(\cos\theta\rho \tilde{w}' v_\theta\right)
=
$$
\begin{equation}
\label{oa1'}
=-\rho\frac{\partial \Phi_g}{\partial R}v_R
+\rho\frac{v_\phi^2 v_R}{R}
-\rho\frac{v_\phi^2 v_\theta}{R}\tan\theta\,,
\end{equation}
where ${\cal E}'$ and $\tilde{w}'$ are the `reduced' (poloidal)
full specific energy and the `reduced' (poloidal) full specific
enthalpy, ${\cal E}'\equiv \epsilon+(v_R^2+v_\theta^2)/2$ and
$\tilde{w}'\equiv\epsilon+p/\rho+ (v_R^2+v_\theta^2)/2$,
respectively.  In all of our calculations equation (2) was used
instead (1e).

\subsection{Characteristic Scales}

The physical quantities have evident characteristic scales:
Distances are measured in terms of the inner radius of the disk
$R_0$.  Velocities are measured in terms of the Keplerian
velocity at the inner edge of the disk, $V_{K0} \equiv
\sqrt{GM/R_0}$ .  Time is measured in units of the period of
revolution of the inner edge of the disk $t_0 \equiv 2\pi
R_0/V_{K0}$.  The density is given in relative units.

\begin{figure*}[t]
\epsscale{0.4}
\plotone{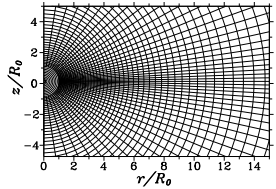}
\caption{
Example of nature of grid used in simulations.  The grid is in
spherical coordinates with $R$ the radial distance and  $\theta$
the co-latitude measured from the equatorial plane.  The grid is
non-homogeneous with increments increasing exponentially with
both $R$ and $|\theta|$.  The grid shown has $N_R\times
N_\theta= 50 \times50$.
}
\end{figure*}

\subsection{Initial Equilibria}

We have adopted spherical coordinates for our simulations in
order to obtain a good match of the grid to the accretion disk
which in general has a thickness in the $z-$direction ($2H$)
increasing with $R$.  However, for convenience our figures and
initial disk equilibria described in this subsection are given
in $(r,z)$ coordinates.

\subsubsection{Initial Conditions 1}

A disk equilibrium with $v_r=0$, $v_z=0$, but $v_\phi \ne 0$ is
described by the equations

$$
\frac{1}{\rho}\frac{\partial p}{\partial r}
=\frac{v_\phi^2}{r}-\frac{\partial\Phi_g}{\partial r}\,,
\eqno(3{\rm a})
$$
$$
\frac{1}{\rho}\frac{\partial p}{\partial z}
=-\frac{\partial\Phi_g}{\partial z}\,,
\eqno(3{\rm b})
$$
where the position and velocity are now dimensionless as
mentioned.

With the assumption of a polytropic gas $p=K\rho^{1+1/n}$, where
$n$ is the polytropic index, $n \equiv 1/(\gamma-1)$, the common
solution of equations (3) can be written as

$$
w=w_0+F-\Phi_g=w_0+F+\frac{1}{\sqrt{r^2+z^2}}\,,\eqno(4)
$$
where $w$ is the specific enthalpy, $w=K(n+1)\rho^{1/n}=nc_s^2$,
with $w_0$ a constant, $c_s = \sqrt{\gamma p/\rho}$ the sound
speed, and $F(r)$  an arbitrary function depending only on $r$.
The angular velocity $v_\phi$ is given in this case as

$$
v_\phi^2=r\frac{\partial F}{\partial r}\,.
$$
The disk boundary is determined by the condition $w=0$ or
$\rho=0$.  If we assume the Keplerian law for angular velocity,
$v_\phi^2=1/r$, then

$$
F=-\frac{1}{r}=\Phi_g(r,z=0)\,,
$$
and

$$
\rho=\rho_0
\left(1-\frac{\frac{\displaystyle 1}{\displaystyle r}
-\frac{\displaystyle 1}{\displaystyle \sqrt{r^2+z^2}}}
{nc_{s0}^2}\right)^n\,,\eqno(5)
$$
where $\rho_0$ and $c_{s0}$ are the values at the point
$(r=1,z=0)$.  The disk boundary in this case is

$$
z^2=\frac{r^2}{1-nc_{s0}^2r^2}-r^2\,,
$$
so the disk thickness diverges at some very large radius outside
the region of interest.  A sample density distribution is shown
in Figure 2.  Hereafter, we refer to this case as Initial
Condition 1.  This was used for the runs described in \S 3 and
\S 4.

\begin{figure*}[t]
\epsscale{0.6}
\plotone{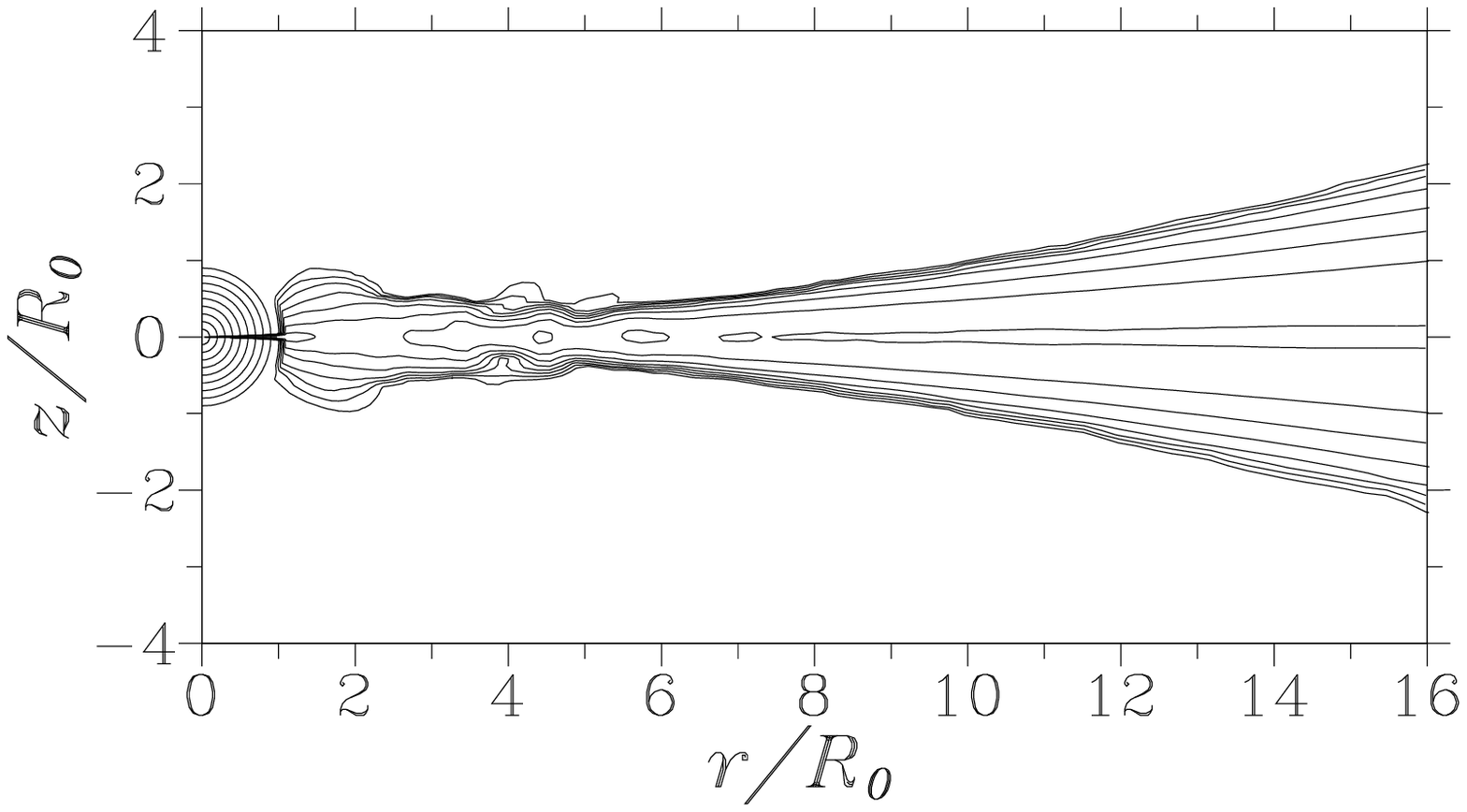}
\caption{{\it Model I.}
Iso-density lines for a conventional disk rotating in one
direction for Initial Conditions 1 after a time $t=40 t_0$,
where $t_0$ is the period of rotation of the inner edge of the
disk at $r=R_0$.  The disk is close to its initial
configuration, excluding the inner part, which has changed as a
result of numerical viscosity.  Simulations were performed with
a grid $100\times100$.
}
\vspace{0.5cm}
\plotone{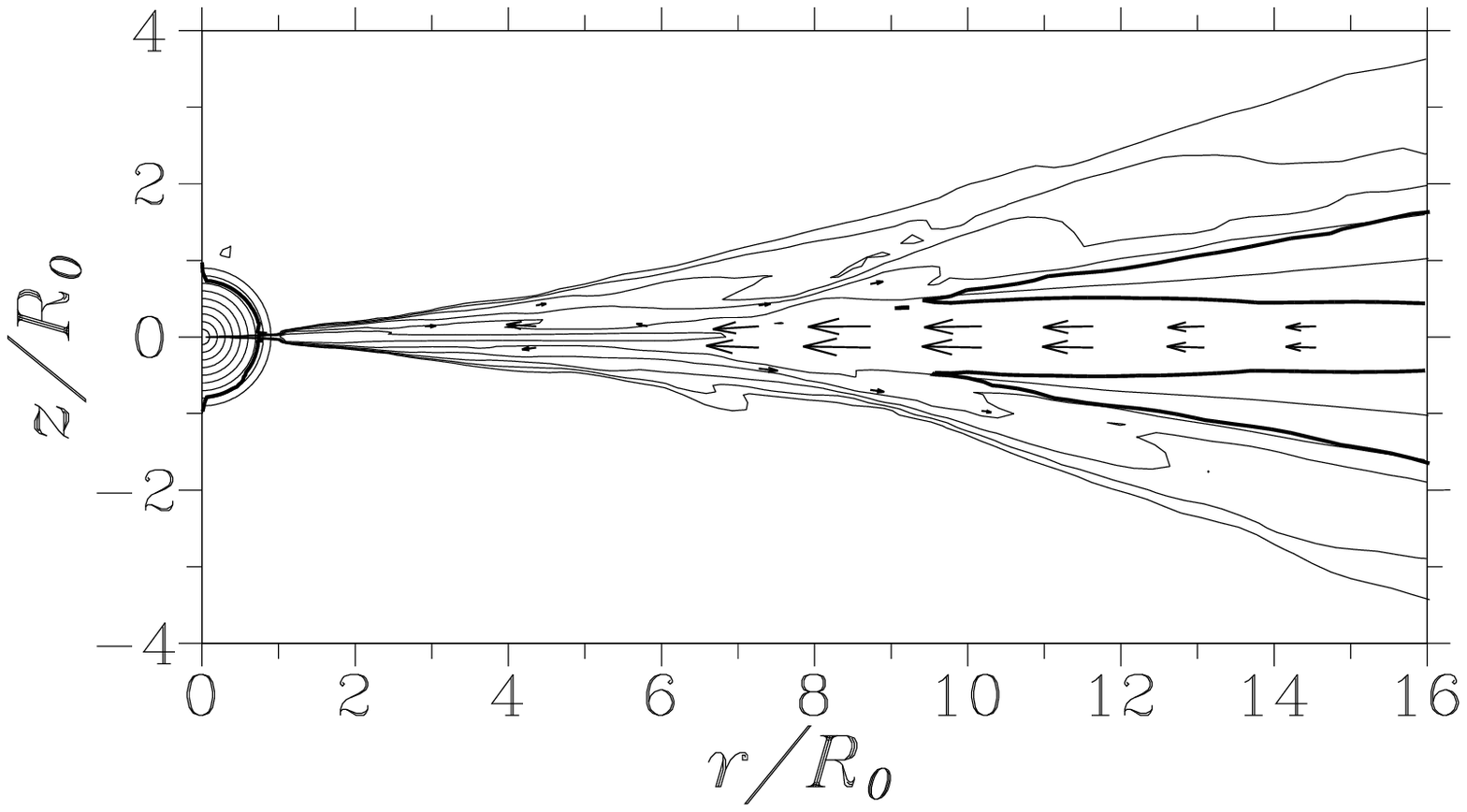}
\caption{{\it Model I.}
Iso-density contours (solid lines) and mass flux vectors
($\rho{\bf v}$) in counterrotating accretion disk at time $t=14
t_0$, where $t_0$ is the rotation period at the inner radius of
the disk. The poloidal flow near the midplane of the disk is
supersonic, excluding the small regions separated by the bold
lines.  Simulations were performed with a grid $100\times 100$.
}
\end{figure*}

\begin{figure*}[t]
\epsscale{0.51}
\plotone{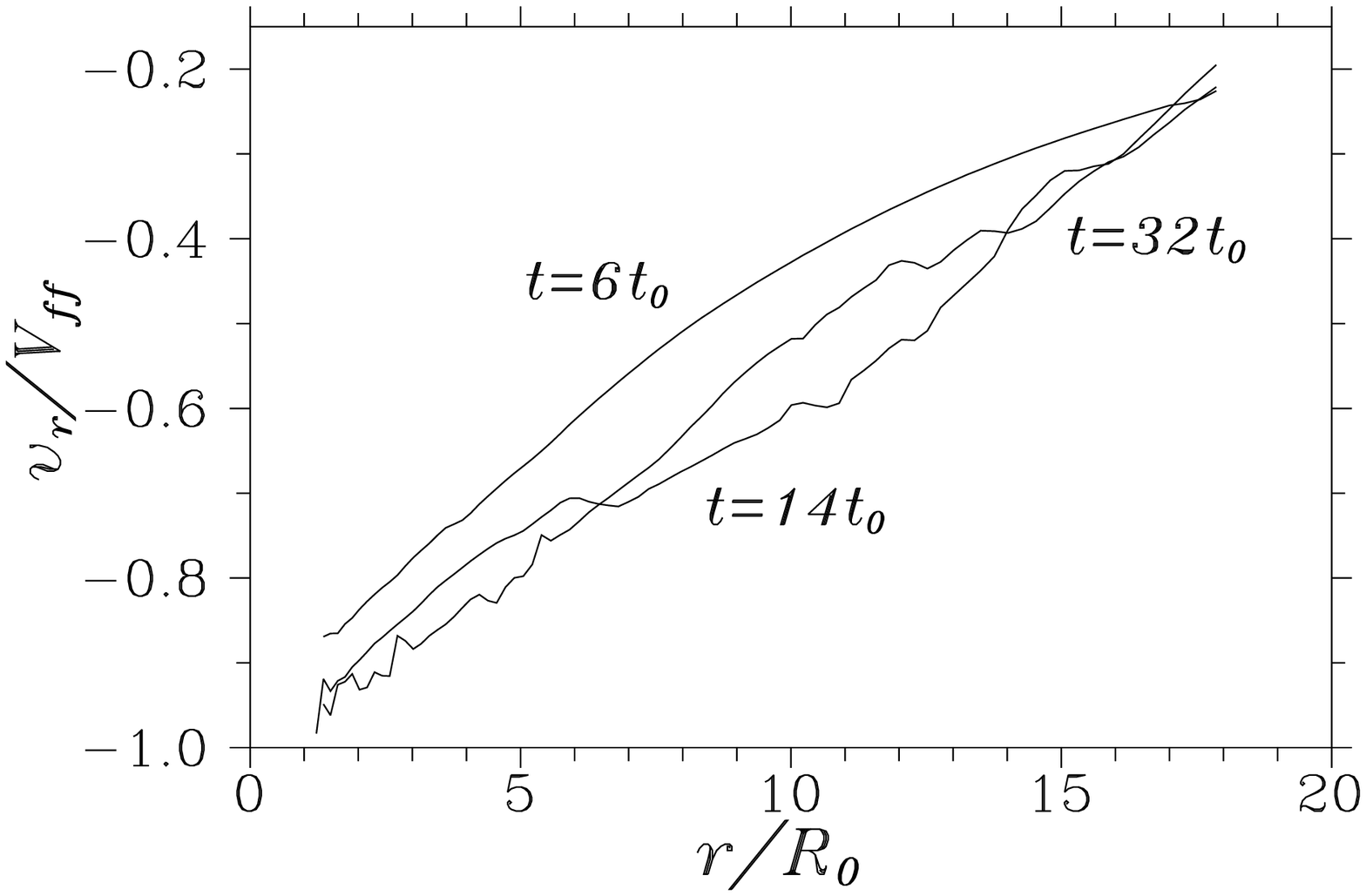}
\caption{{\it Model I.}
Variation of radial velocity $v_r$ with $r$ in the midplane of
the disk at times $t=6 t_0$, $14t_0$, and $32t_0$,
where $t_0$ is the rotation period at the inner radius of the
disk.
}
\vspace{0.5cm}
\plotone{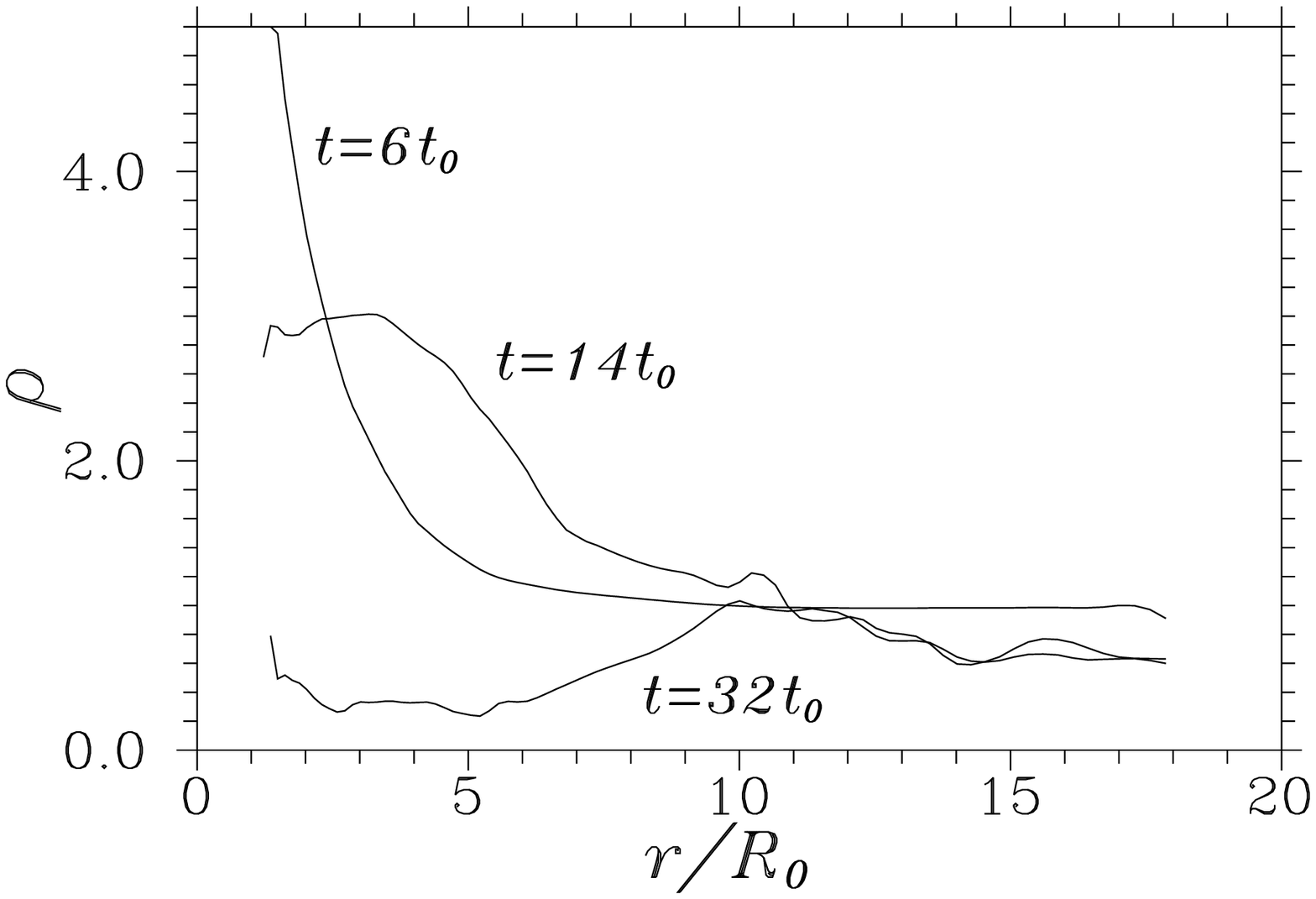}
\caption{{\it Model I.}
Variation of density with $r$ in the midplane of the disk at the
same time intervals as Figure 7.
}
\end{figure*}

\subsubsection{Initial Conditions 2}

A second equilibrium disk model was used in our simulations.
For this we took the disk boundary ($\rho=0$) to have the form

$$
z^2=
r^2\left(1-\frac{r^2}{r_1^2}\right)\Delta^2\,,
$$
where $r_1$ and $\Delta$ are parameter determining disk radial
size and thickness.  The rotation (slightly sub-Keplerian),
enthalpy, and density for this case are:

$$
v_\phi^2=\frac{1}{r\sqrt{1+\Delta^2}}
\frac{1-\frac{\displaystyle 2r^2}
{\displaystyle r_1^2}\beta}
{\left(1-\frac{\displaystyle r^2}{\displaystyle
r_1^2}\beta\right)^{3/2}}\,,\eqno(6{\rm a})
$$

$$
w=\frac{1}{\sqrt{r^2+z^2}}
-\frac{1}{r\sqrt{1+\Delta^2-
\frac{\displaystyle r^2}{\displaystyle
r_1^2}\Delta^2}}\,,\eqno(6{\rm b})
$$

$$
\rho=\rho_0 w^n\,, \qquad \beta=\frac{\Delta^2}{1+\Delta^2}\,.
\eqno(6{\rm c})
$$
The density distribution for this case is shown in Figure 3.
Hereafter, we refer to this case as Initial Conditions 2.

For building equilibria we have used both a rotation law based
approach and one based on the shape of the disk surface.  For
the first case, the equilibria are found for known rotation law
according to the chain $v_\phi(r) \mapsto F(r) \mapsto w(r,z)
\mapsto \left.z(r)\right|_{\rho=0}$.  For the second case, a
known disk surface is used according to the chain
$\left.z(r)\right|_{\rho=0} \mapsto F(r)=\Phi(r,z(r))-w_0
\mapsto v_\phi(r)$.  Abakumov et al. (1996) use an analogous
approach.

\subsection{Boundary Conditions}

Simulations were performed in the region $R_0 \leq R \leq
R_{max}$, $-{\pi/ 2} \leq \theta \leq {\pi/ 2}$.

The outer boundary conditions $R=R_{max}$ are treated following
the method used by Sawada based on solving of the Riemann
problem (see, e.g., Sawada, Matsuda, \& Hachisu 1986; Sawada \&
Matsuda 1992).  These boundary conditions were used for Models I
and II.  However, for Model III different boundary conditions
are applied in that matter is considered to inflow
supersonically through part of the outer boundary.

We assume that all matter coming into the sphere $R \leq R_0$ is
absorbed.  On the $z-$axis, $\theta=\pm {\pi/ 2}$, we of course
have $v_\theta=0$ and $v_\phi=0$.

No symmetry is assumed with respect to the equatorial plane.

\subsection{Numerical Method and Viscosity Treatment}

An explicit Lax-Friedrichs finite-difference scheme with flux
correction in Chakravarthy-Osher (Chakravarthy \& Osher 1985)
form was used.  This scheme is $1st$ order of approximation in
time and $3rd$ order of approximation in space and is
oscillation-free near discontinuities.  The combined
Lax-Friedrichs-Osher scheme was suggested in work by Vyaznikov,
Tishkin, \& Favorsky (1989).

Figure 4 shows the non-homogeneous $(R,\theta)$ grid used in the
simulations.  The radial grid increment $\delta R$ increases
outwards exponentially, $\delta R = \delta R_0 a^{R/R_0}$, where
$a=1.05$.  The $\theta$ grid $\delta \theta$ increases in a
similar way going away from the equatorial plane.  For a disk
which has thickness $h$ increasing with $r$, this type of grid
gives a more uniform distribution of cells within the disk
including its central regions.  In the simulations we used grids
of  $N_R \times N_\theta$ of $100 \times 100$ or $200 \times
200$ in some cases.

The simulation method used naturally has numerical viscosity.
With a fixed grid, we cannot vary the numerical viscosity, but
we can investigate its influence.  The coefficient of numerical
viscosity can be represented as $\nu_{num} \sim  c_s \Delta h$,
where $c_s$ is the sound speed, $\Delta h$  is the size of the
grid.  This expression for $\nu$ is analogous to that proposed
by Shakura (1973) and Shakura \& Sunyaev (1973) for the
turbulent viscosity of accretion disks,  $\nu=\alpha c_s H$,
where $H$ is the thickness of the disk, and $\alpha\sim 10^{-1}
- 10^{-3}$ is a dimensionless parameter.  We calculated the
steady accretion for a disk rotating in one direction and
observed the accretion due to the numerical viscosity.  Using
the formula for the accretion speed of a viscous accretion disk,
$v_r \sim \nu/r$, we estimated that $\alpha < 1$ for $r>5 R_0$
for a $200\times 200$ grid and $\alpha < 1$ for $r>7 R_0$ for a
$100\times 100$ grid, and $\alpha$  decreases  to $\alpha\sim
0.03 - 0.1$ at the outer radii.  A challenge for future work is
the inclusion of a true $\alpha$ viscosity.

We investigated three different models of counterrotating
accretion flows which correspond to different possible
astrophysical situations.  The first model corresponds to disks
counterrotating in the $z-$direction (Model I, \S 3).  The
second corresponds to disks counterrotating in the $r-$direction
(Model II, \S 4).  The third corresponds to counterrotating gas
infalling onto a pre-existing corotating disk (Model III, \S 5).

\begin{figure*}[t]
\epsscale{0.5}
\plotone{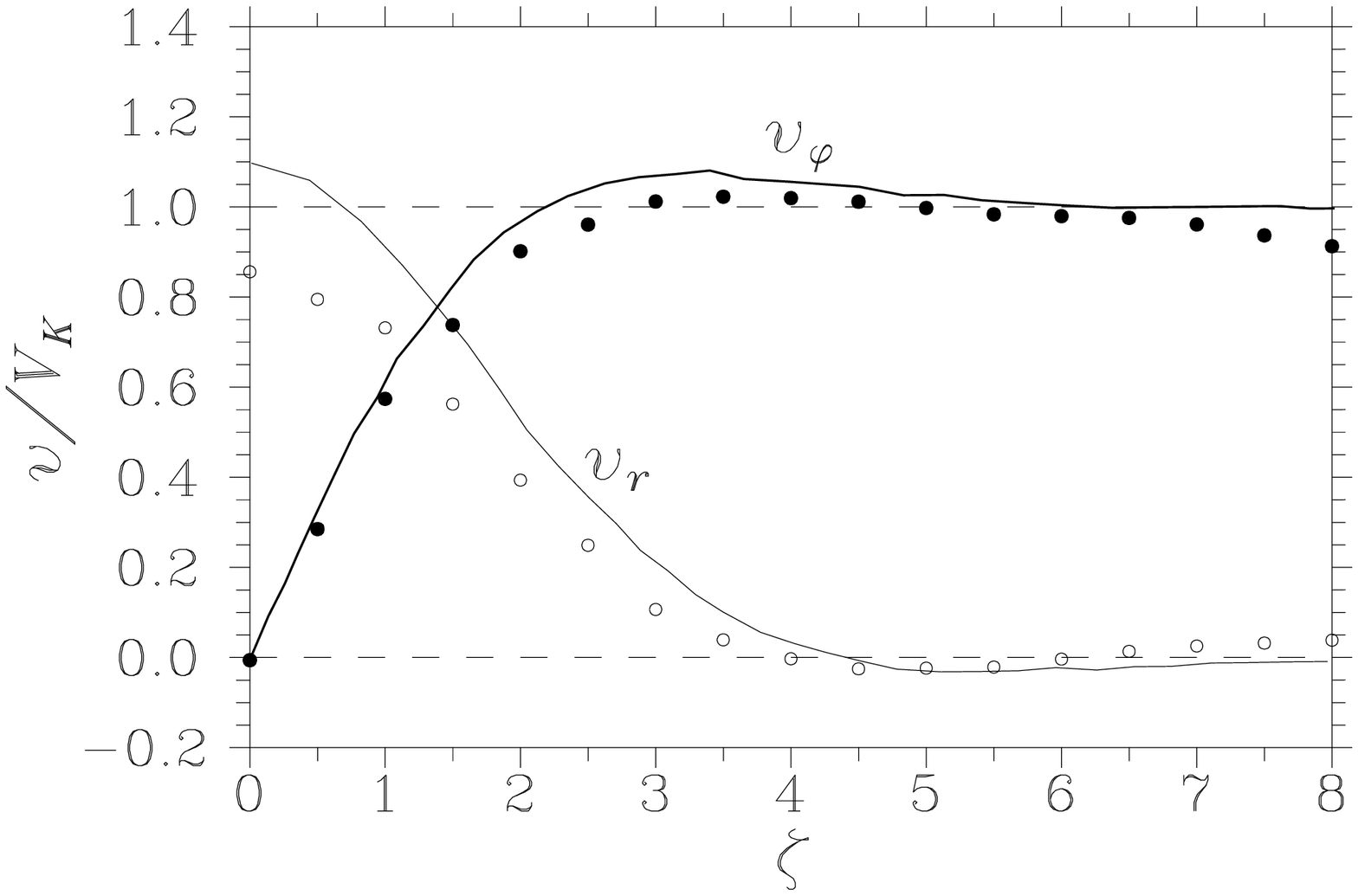}
\caption{{\it Model I.}
Radial velocity $v_r$ (hollow circles) and azimuthal velocity
$v_\phi$ (filled circles) velocities across the disk as a
function of $\zeta \equiv z/h$ at radial distance $r=7R_0$.
Here, $h(r)$ is the characteristic half-thickness of the region
of the strongest inflow, and $V^2_K \equiv GM/r$.  The
solid lines show the theoretical dependencies from LC.  Note that
to a good approximation $v_r$ is an even function of $\zeta$ and
$v_\phi$ is an odd function.
}
\plotone{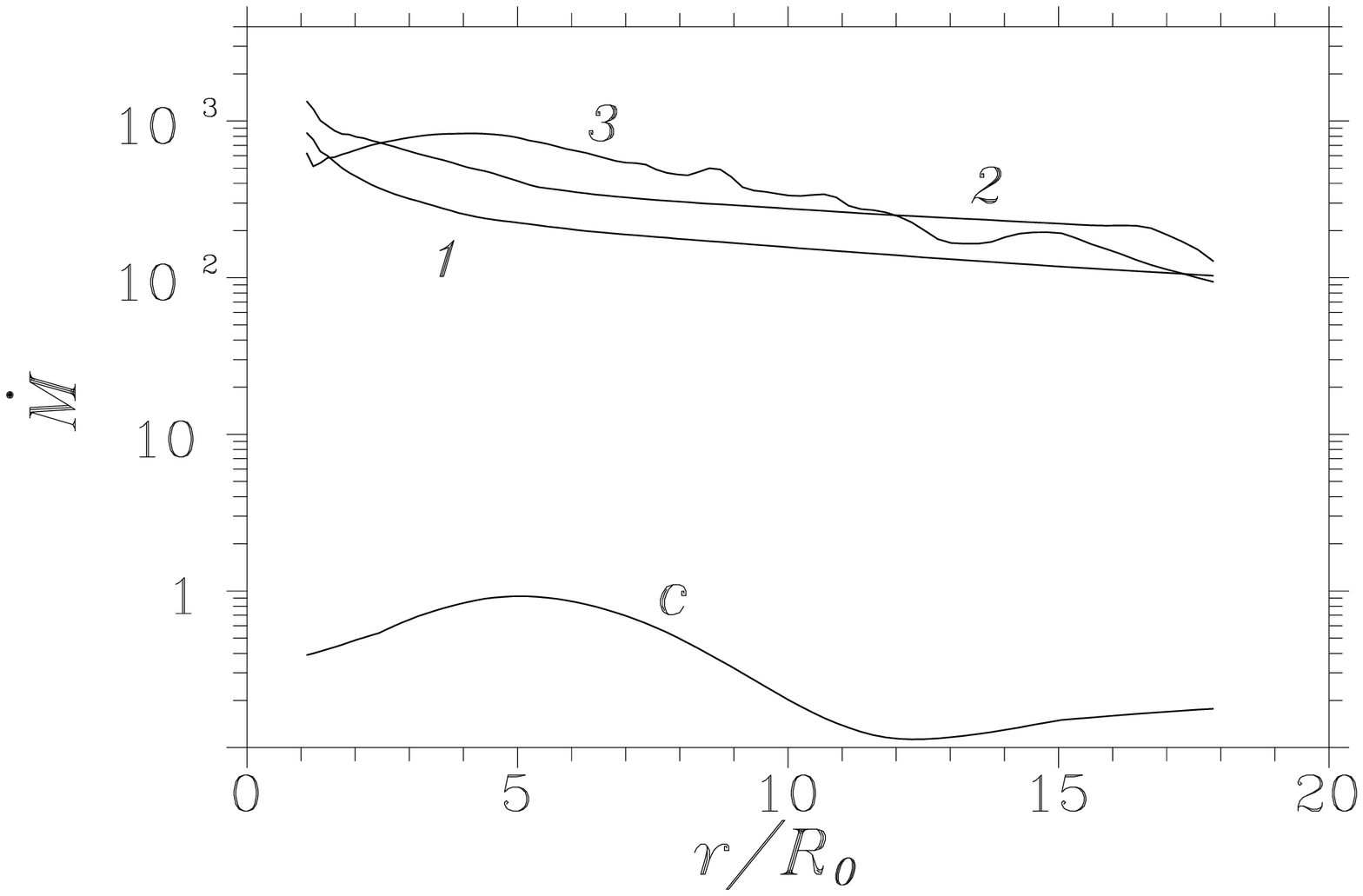}
\caption{{\it Model I.}
Mass accretion rate as a function of radial distance $r$ at
different moments of time.  The upper curves are for a
counterrotating disk at times $t=3t_0$, $7t_0$, and $14t_0$. The
lower curve (C) is for a conventional accretion disk rotating in
one direction at time $t=40t_0$, where the accretion results
from the numerical viscosity of the code.
}
\end{figure*}

\section{Disk Counterrotating in the $z-$direction: Model I}

Here, we discuss simulation results for Model I where initially
the upper half of the disk ($z>0$) is corotating ($+\Omega_K$)
while the lower half ($z<0$) is counterrotating ($-\Omega_K$).
This case was treated analytically by LC, and therefore we have
a test both of the LC theory and of our simulation code.

First in \S 3.1, we check the initial equilibrium disk for the
conventional case of unidirectional disk rotation.  Secondly in
\S 3.2, we investigate the counterrotating disk for
$\gamma=1.01$, which is close to isothermal.  This value of
$\gamma$ close to unity mimics conditions of strong radiative
cooling.  In this subsection we compare the simulations results
with the predictions of LC where radiative cooling is included.
In these subsection we used a $100\times 100$ grid.  Thirdly in
\S 3.3, we discuss the cases $\gamma=1.1$ and $5/3$ which
corresponds to non-isothermal conditions where numerical viscous
heating is significant. In this subsection we used a $200\times
200$ grid.

\begin{figure*}[t]
\epsscale{1.0}
\plotone{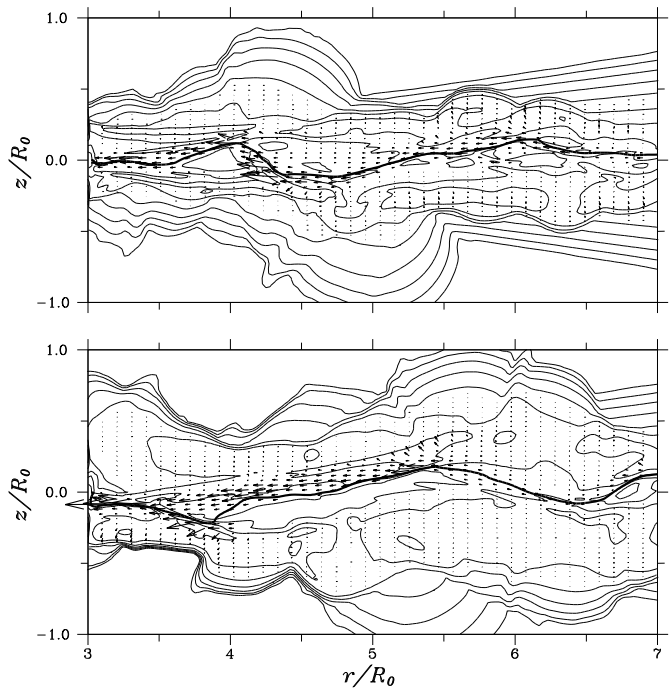}
\caption{{\it Model I.}
Contours of density and velocity vectors for $\gamma=1.1$ at
times $t=4.8 t_0$ (bottom panel) and $t=6.2 t_0$ (top panel).
The bold line separates the corotating and counterrotating gas.
The simulations were performed for the part of the disk $3  \leq
r/R_0 \leq 7 $ with a $200 \times 200$ grid.
}
\end{figure*}

\begin{figure*}[t]
\epsscale{0.5}
\plotone{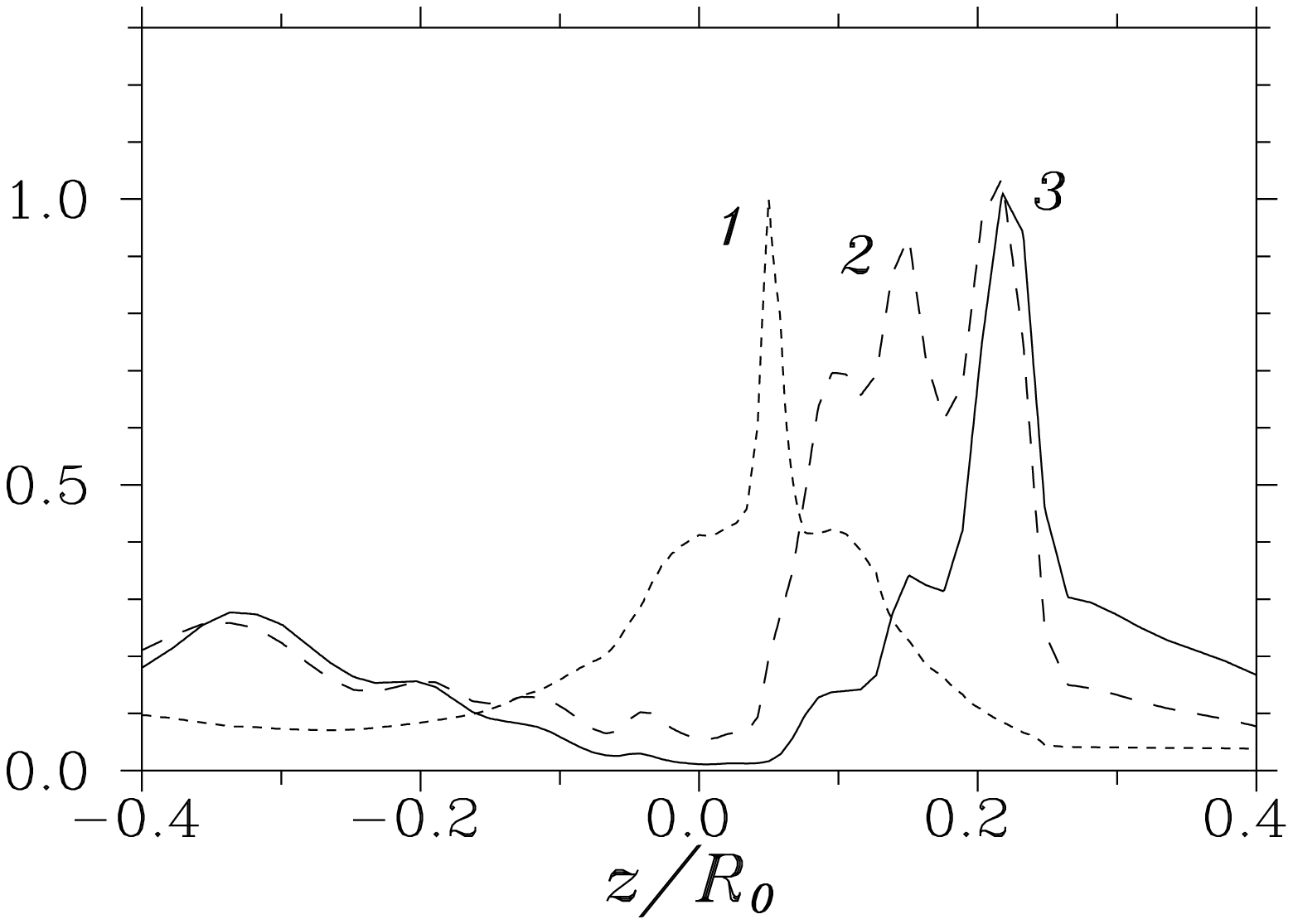}
\caption{{\it Model I.}
Profiles of the density (1), pressure (2), and temperature (3)
through the disk at $r= 5R_0$ at the time $t=6.2t_0$
corresponding to the top panel of Figure 11.
}
\end{figure*}

\subsection{Conventional Accretion Disk}

As a calibration of our simulation code, we investigated a
conventional accretion disk which rotates in only one direction.
The disk evolves as a result of the relatively small numerical
viscosity of the code.  We took Initial Conditions 1 (see \S
2.4.1) with $\gamma=1.01$ and performed long-term numerical
simulations of this disk.  We observed that the disk shape did
not change during many periods of rotation of the inner radius
of the disk.  Figure 5 shows the iso-density contours
distribution at $t=40 t_0$.  The evolved disk is close to that
at $t=0$ (see Figure 2), excluding the inner part, where the
thickness of the disk is very small (comparable with the grid
size), and the numerical viscosity has a noticeable influence.

\subsection{$\gamma=1.01$}

This case is close to isothermal and is closest to the LC
analytical model where both viscous heating and radiative
cooling are included.

We took Initial Conditions 1 for the initial conditions for the
density and velocity of the disk (see Figure 2), but put the
angular velocity in the lower-half of the disk ($z<0$) opposite
to the angular velocity of the upper-half ($z > 0$).  Figure 6
shows that matter starts to rapidly flow radially inward in a
narrow region near the midplane of the disk.  The poloidal flow
near the midplane is supersonic, and the radial inflow velocity
increases as $r$ decreases.

Figure 7 shows the radial variation of the radial velocity in
the midplane of the disk at different times.  The velocity
distribution changes only gradually with time.  The velocity
close to the gravitating center, $r/R_0 \sim 1-3$ is much larger
than velocity in the outer regions of the disk, $r/R_0 \sim
16-18$.  Note, that for $r \sim R_0$,  the radial inflow
velocity is close to the free-fall velocity
$V_{ff}=\sqrt{2GM/R}$.  An inflow velocity $\sim \sqrt{GM/R}$ is
predicted by LC.  Velocity is strongly supersonic with Mach
number $M>100$. Figure 7 shows that velocity distribution along
the disk is almost the same during the long time.  Figure 8
shows the dependence of density on radial distance.

The density of the disk decreases with time, specifically in the
inner regions of the disk.  Both, the velocity and density show
small-scale wave-like variations as a function of radius.

Figure 9 shows the observed vertical ($z$) variation of the
radial and azimuthal velocities in the disk at $r=7 R_0$ and the
theoretical dependencies calculated by LC.  The dimensionless
vertical length $\zeta$ was introduced by LC  as $\zeta=z/h$,
where $h=h(r)$ is the characteristic half-thickness of the
boundary layer in which most of the radial inflow of matter
occurs.  Consistent with LC, we find that $(h/H)^2 \ll 1$ over
most of the disk surface, where $H$ is the full disk
half-thickness.  At $r=7R_0$, we find $h/H \approx 0.25$.  One
can see that radial inflow velocity has a maximum in the
midplane of the disk and decreases to small values at large
$\zeta$. Note that we observe weak outflows for $4 < \zeta < 8$
where $v_r > 0$.  Similar outflows were predicted by LC (the
solid curves).  The azimuthal velocity $v_\phi$ grows with
increasing $\zeta$ and reaches the Keplerian velocity for
$\zeta~ {\buildrel > \over \sim}~ 2.4$.  There is also a region,
where the azimuthal velocity is slightly larger than Keplerian
velocity, $3 < \zeta< 4.5$.  A similar region was predicted by
LC (see solid curve in Figure 9).

The conditions considered by LC were different from the initial
conditions considered for our simulations.  Thus, it appears
that the main features of vertically stratified counterrotating
flows do not depend strongly on the initial conditions.  At
longer times in our simulation runs, the region of fast
equatorial flow becomes somewhat thicker.  Overall the
simulation results are close to the predictions of the analytic
model of LC.  Although the analytic model includes viscous
heating and radiative cooling, these almost compensate each
other with the result the  model is close to the simulations
which has almost isothermal conditions and no radiative cooling.

Figure 10 shows the radial dependence of the mass accretion rate
$\dot M(r)$ at different times, and that for a disk rotating in
one direction at time $t=40t_0$ (\S 3.1).  The mass accretion
rate for the counterrotating disk can be written as

$$
\dot M \approx 2\pi r \rho(r,z=0) V_K(r)[3.25 h(r)]\,,\eqno(7)
$$
where the numerical coefficient (3.25) is obtained from Figure
9.  In that the surface mass density of the disk is $\Sigma =
2\pi \rho(r,z=0)2H$, the average {\it accretion speed} is
$u_{CR} \approx 1.62(h/H)V_K$.  For comparison the mass
accretion rate for a conventional $\alpha-$disk rotating in one
direction (Shakura 1973; Shakura \& Sunyaev 1973) is

$$
\dot M_{SS} \approx 2\pi r \rho(r,z=0)V_K(r)
\left({\alpha c_s^2 \over V_K^2}\right) 2H(r)\,.
\eqno(8)
$$
In this case the average accretion speed is $u_{SS} = \alpha
(c_s/V_K)^2V_K$.  For a conventional accretion disk, the
quantity $\alpha (c_s/V_K)^2$  is commonly thought to be much
smaller than unity.  For example, for $c_s/V_K = 0.01$, $h/H =
0.1$, and $\alpha = 0.1$, $\dot M/\dot M_{SS} \approx 1.6 \times
10^4$.

\subsection{$\gamma=1.1$ and $\gamma=5/3$}

We also did simulations of vertically stratified counterrotating
disks for $\gamma=1.1$ and $\gamma=5/3$ which correspond to
non-isothermal conditions were the affect of viscous heating is
more important.  We find in general that after a given time and
for a given numerical grid, the temperatures in the disk
increase as $\gamma-1$ increases.  Thus, the disk thickness
$H/r$ is larger for larger $\gamma-1$.  On the other hand for
fixed $\gamma$, the temperature in the disk decreases gradually
as the resolution $N_R \times N_\theta$ increases.  Here, we
performed simulations with the grid $200\times 200$. To increase
the resolution even more (to decrease viscosity) we performed
simulations of only part of the disk: $3 R_0<r<7 R_0$.

At early times ($t/t_0 \sim 1-3$), the flow in the equatorial
plane of the disk was similar to that for $\gamma=1.01$.
However,  for longer times the region of fast poloidal inflow
begins to warp.  Figure 11 shows contours of density and
velocity vectors of the disk at two times.  The amplitude of the
warp of the interface $\Delta z$ between corotating and
counterrotating flows grows.  After a fixed time the warp
amplitude is a strongly increasing function of $\gamma-1$.  For
$\gamma = 1.01$ the fractional warp amplitude $\Delta z/r$ is
very small for all times.  However, the warp amplitude for
$\gamma=1.1$ is $\Delta z/r \approx 0.034$, and for $\gamma=5/3$
it is $\Delta z/r \approx 0.13$ both at $t/t_0 =7$ with the same
resolution as used in Figure 11.  The warp amplitude $\Delta z$
is of the order of the disk half-thickness $H$.  For $\gamma =
1.1$ the dominant radial wavelength is $\lambda_r \sim 1.5R_0$
while for $\gamma=5/3$ there is a wide spectrum of wavelengths
from $\sim R_0$ to $\ll R_0$.  The warping is suppressed by low
grid resolution.

The warping may represent a Kelvin-Helmholtz type of instability
driven by the shear in the poloidal flow.   The tidal
gravitational force of the central object, $F_z$, acts to keep
the warp amplitude from growing without bound.  Note that
Kelvin-Helmholtz instabilities driven by the shear in the
azimuthal flow are suppressed in our simulations which are
axisymmetric.  Alternatively, the disk warping may be due to
unsymmetrical (top/bottom) heating of the disk which gives an
excess pressure in say the bottom half of the disk which
displaces the disk upward.  Plots of the pressure, density, and
temperature through the disk as shown in Figure 12 support this
simple explanation.

\begin{figure*}[p]
\epsscale{0.44}
\plotone{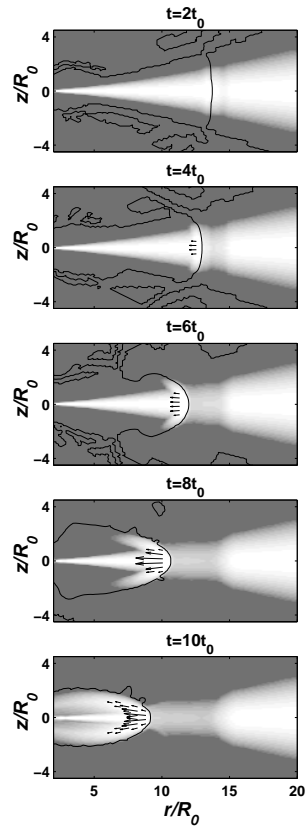}
\caption{{\it Model II.}
Gray-scale plots of density and matter flux vectors showing the
evolution of Model II.  Starting from the top, the panels show
times $t=2 t_0$, $4 t_0$, $6 t_0$, $8 t_0$, and $10 t_0$.  The
vectors are proportional to the value of the mass flux $\rho{\bf
v}$.  The black solid line shows the line of zero azimuthal
velocity.  For the case shown, $\gamma=1.1$.
}
\end{figure*}

\begin{figure*}[t]
\epsscale{1.0}
\plotone{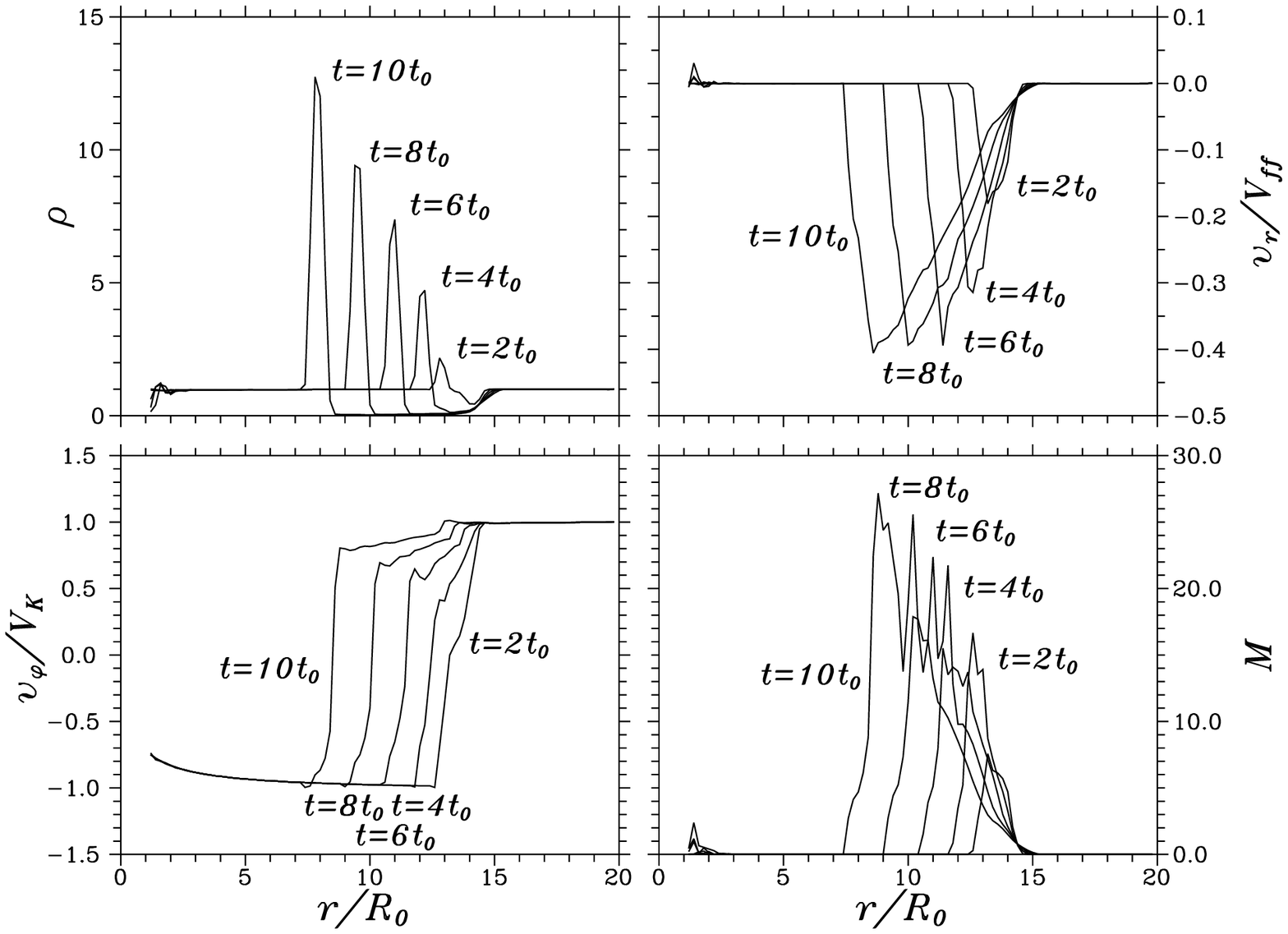}
\caption{{\it Model II.}
Radial distributions of density (top left panel), radial
velocity (in units of local free-fall velocity) (top right
panel), azimuthal velocity (in units of local Keplerian
velocity) (bottom left panel), and Mach number (right bottom
panel) at different times, $t=2 t_0$, $4 t_0$, $6 t_0$, $8
t_0$, and $10 t_0$.  The time sequence corresponds to
that shown at Figure 13.
}
\end{figure*}

\section{Disk Counterrotating in the $r-$Direction:  Model II}

Here, we treat Model II which is shown schematically in the
middle panel of Figure 1.  In this model the inner part of the
Keplerian disk ($r < r_t$) rotates in the positive direction
whereas the outer part ($r > r_t$) rotates in the opposite
direction.  The transition between positive and negative
azimuthal velocity was centered at $r/R_0=14$ and extended over
the narrow region $13.5 < r/R_0 < 14.5$.  This transition region
is small, involving only $5$ grid points in the radial direction
for a mesh  size of $100\times 100$.  The boundary condition at
the right side of the disk was taken the same as in Model I.
Simulations were performed for $\gamma=1.1$ and $\gamma=5/3$.
Results are presented for $\gamma=1.1$.

\begin{figure*}[t]
\epsscale{0.7}
\plotone{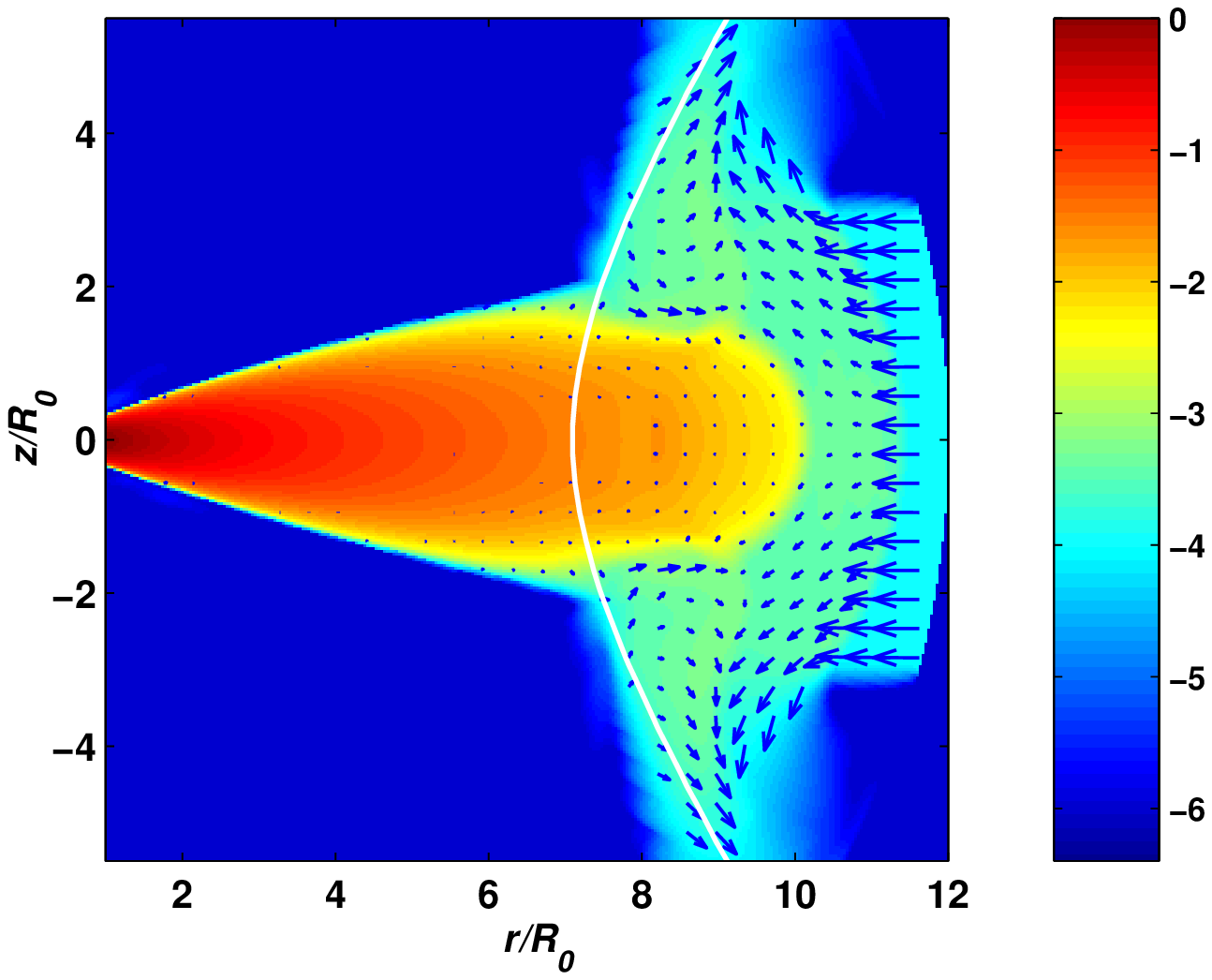}
\caption{{\it Model III.}
Results of simulations of Model III where radially inflowing
matter encounters an existing corotating disk.  The inflowing
matter rotates in the same direction as the disk and there is no
accretion.  The inflowing matter is injected at $z \sim \pm 3
R_0$.  The logarithm of the density is represented by the color
scale shown.  The white line shows centrifugal barrier for the
inflowing gas.  The vectors  show the poloidal velocity.  The
plot is  for $t= 20 t_0$. Simulations were performed with
resolution $N_R=150$ and $N_\theta=100$.
}
\end{figure*}

In our simulations,  matter in the transition zone with low
angular momentum flows rapidly inward compressing matter at
smaller radii as shown in Figures 13 and 14.  This inflow sets
up a density wave which propagates inward in the disk.  The
matter behind the wave has lower density.  The speed of
propagation of the wave grows with time and reaches a maximum
value equal to about one half of the free-fall velocity (see
Figure 14).  The azimuthal velocity in the wave is sub-Keplerian
with $v_\phi \sim (0.6-0.8) v_K$ in inner part of the disk.
Matter behind the wave (at larger $r$) has opposite  velocity,
with a magnitude somewhat smaller than the local Keplerian
velocity (see Figure 14).  The wave propagates supersonically,
with Mach number $M\sim 20-30$.  The Mach number grows as the
wave moves inward (see Figure 14).

\begin{figure*}[p]
\epsscale{0.7}
\plotone{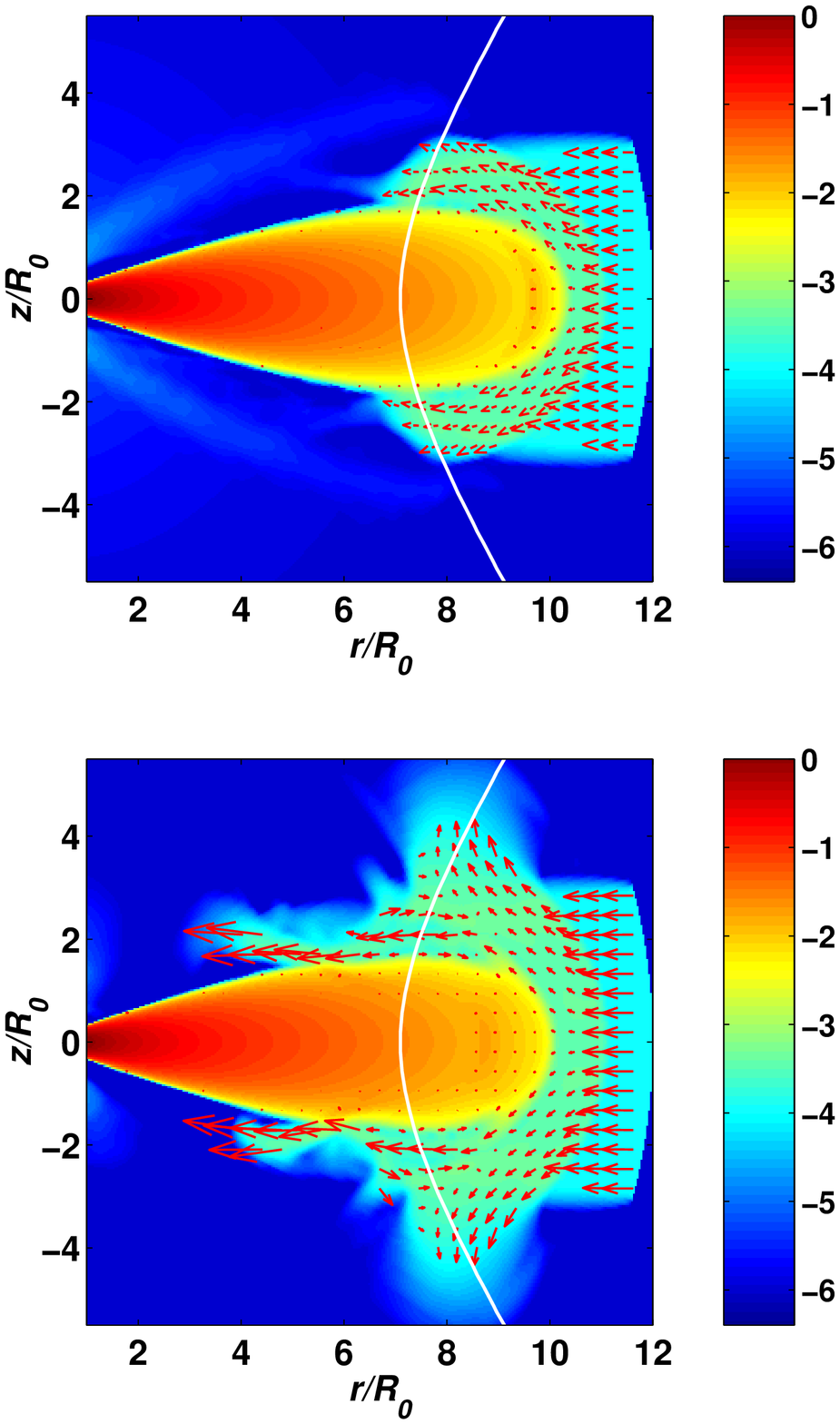}
\caption{{\it Model III.}
Results of simulations of Model III for the case where the
inflowing matter rotates in the {\it opposite} direction to that
of the disk.  In this case, incoming counterrotating matter
interacts strongly with the main disk.  The resulting low
angular momentum gas starts to accrete rapidly along the
surfaces of the disk.  The times  $t=4.5 t_0$ (top panel) and $7
t_0$ (bottom panel) of evolution are shown.  The logarithm of
the density is represented by the color scale shown.  The white
lines in both panels show centrifugal barrier for the inflowing
gas.  The vectors  show the poloidal velocity.
}
\end{figure*}

\begin{figure*}[p]
\epsscale{0.7}
\plotone{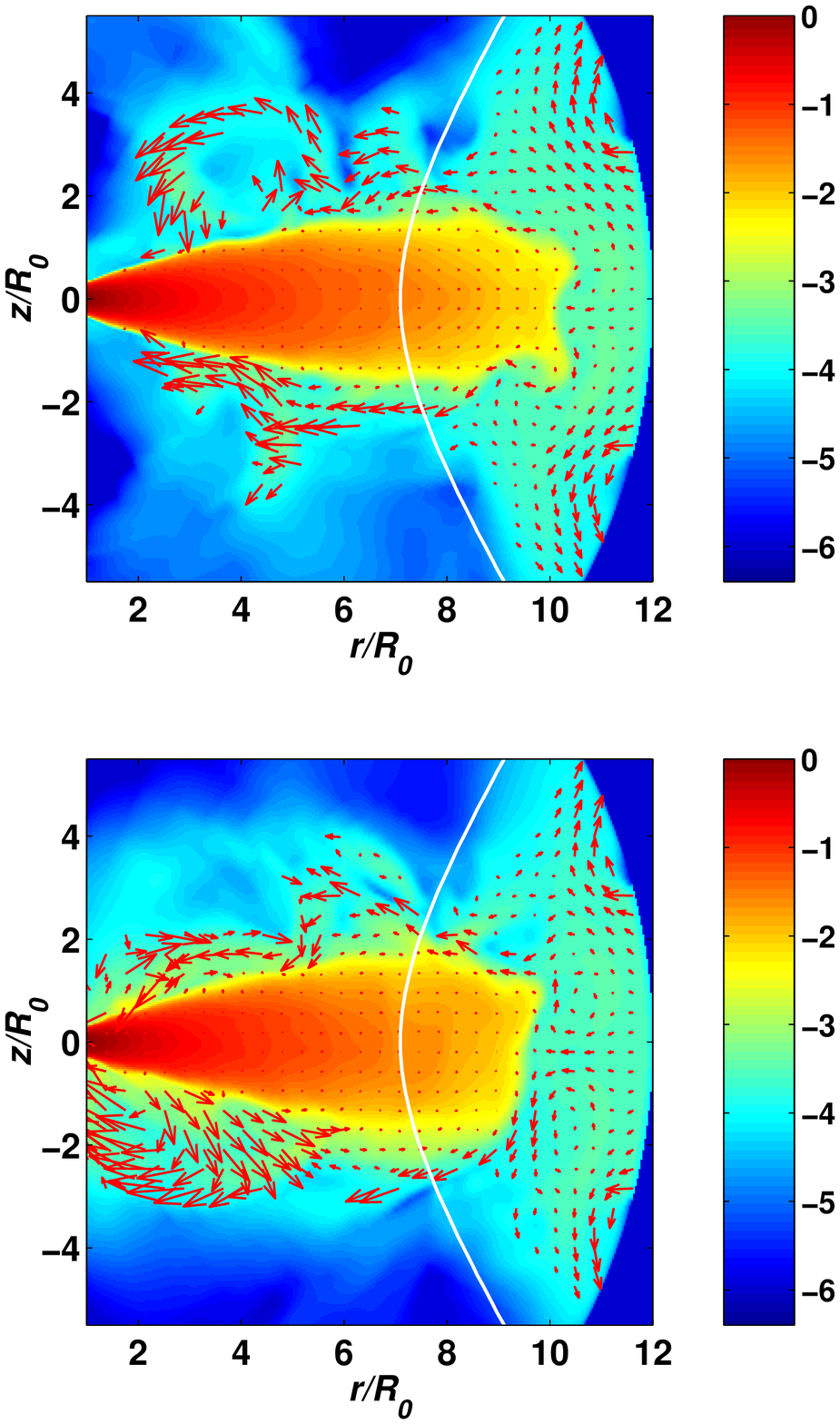}
\caption{{\it Model III.}
Continuation of Figure 16. The top panel is for  $t=20 t_0$,
when the  surface layer accretion is strong.  The bottom panel
is for $t=35 t_0$, when the surface accretion is destroyed by
the dissipative heating.
}
\end{figure*}

The shock wave propagating initially at times $t<10 t_0$ is not
a standard shock wave, because it has significant angular
momentum, and its formation and propagation is based on the
deficit of angular momentum of the matter of the wave and an
inbalance of gravitational and centrifugal forces.  The density
and velocity jumps are much larger than in the case of standard
shock wave (see Figure 14).  This wave resembles a soliton-type
wave.  A similar type of wave was observed in accretion disks
where a local deficit of angular momentum  was caused by
magnetically driven outflows (Lovelace, Romanova, \& Newman
1994; Lovelace, Newman, \& Romanova 1997).  Note that the solid
line in Figure 13 indicates the boundary between corotating and
counterrotating components.  The density wave reaches $r=8 R_0$
during $t=10 t_0$ (ten periods of rotation of the inner edge of
the disk).  For longer times, matter started to ``reflect'' from
the inner regions and move outward.

The effective collapse of the inner disk can be understood by
considering the details of the model used.  The density of the
disk is constant along the equatorial plane so that the mass of
the equidistant rings $\Delta R={\rm const}$ increases outwards,
$\Delta M=2\pi\rho r \Delta r \sim r$, so that the mass of ring
at $r=14$ is $14$ times larger than a ring at $r=1$. At the same
time the angular momentum of the rings increases outwards as
$\Delta L \sim \Delta M v_K r \sim r^{3/2}$, so that the angular
momentum of inner ring is $52$ times smaller than of the ring at
$r=14$.  The layer between counterrotating disks has pretty large
mass and zero angular momentum. It is not centrifugally
supported and falls down to the center.  It accumulates the
matter of the inner regions of the disk and their angular
momentum. But absorbed angular momentum is not sufficient to
rich the centrifugal barrier.

The thickness of the boundary layer in which there is
annihilation of angular momentum depends on the viscosity.  For
a larger viscosity, the layer is thicker and the propagating
wave is stronger.  The fast inward propagation of such a wave in
a disk around a black hole would give an outburst in the
luminosity as the  wave reaches the black hole.  This represents
a possible mechanism of generating outbursts in Active Galactic
Nuclei, for example.

\section{Inflow of Counterrotating Gas onto a Corotating Disk:
Model III}

Here, we discuss Model III which is shown schematically in the
bottom panel of Figure 1.  In this case counterrotating gas of
relatively low density inflows with significant poloidal speed
and encounters an existing Keplerian disk.  For the simulations,
we took Initial Conditions 2 (\S 2.4.2) corresponding to a
finite radius, slightly sub-Keplerian, corotating disk (see
Figure 3).  The effective outer radius of this disk was taken to
be at $r_{out} =11.5R_0$.  At the outer, right-hand boundary of
the computational region, matter is launched into the simulation
region from the part  of the boundary $-3 < z/R_0 < 3 $ with
radial velocity $v_r = 0.7 V_K$ and azimuthal velocity $v_\phi =
-V_K$.  When both: the ``target" disk and incoming matter rotate
in the same direction, then incoming matter moves inward only
until it reaches the centrifugal barrier (see white line on
Figure 15), then it stops, and moves in the $\pm z$ directions
away from the disk surface.

\begin{figure*}[t]
\epsscale{0.55}
\plotone{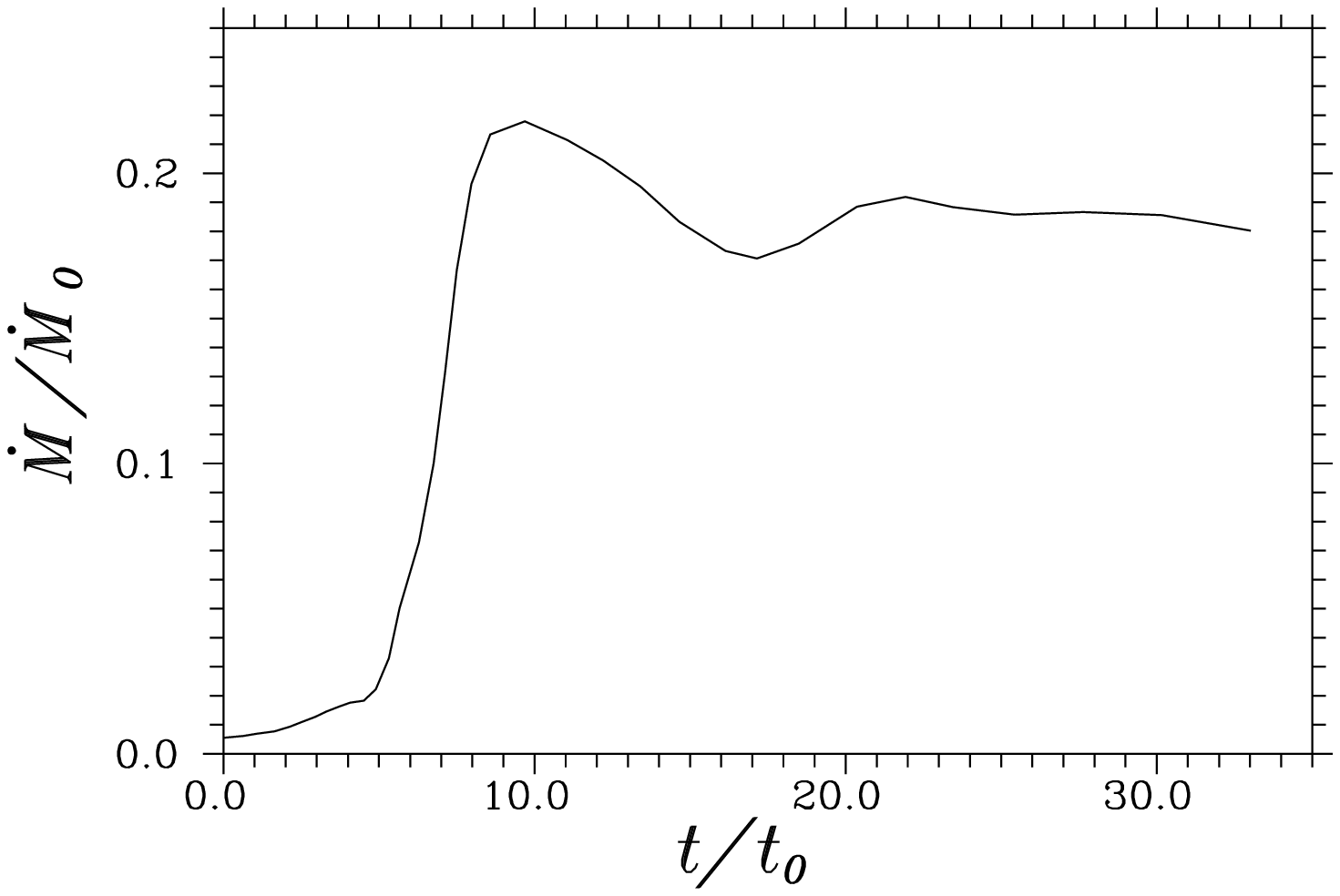}
\caption{{\it Model III.}
Matter flux through the cylindrical surface $r=5 R_0$ in Model
III. It has a peak owing to the surface accretion, then is
almost constant owing to both surface accretion and enhanced
accretion inside the main disk.
}
\end{figure*}

Figure 15 shows the observed behavior of density and velocity in
this case.  Note, that in this case we observe only very small
accretion connected with finite viscosity of the disk. This case
was considered for reference.

When the inflowing matter rotates in the {\it opposite}
direction to that of the ``target'' disk, then it also reaches a
centrifugal barrier.  However, a non-negligible fraction $f$ of
the inflowing matter interacts with the disk causing angular
momentum annihilation.  For this case the fraction is $f \sim
0.2$.  Figures 16 and 17 show the evolution.  One can see from
Figure 16 (top and bottom panels), that at times $3 < t/t_0 <
7$, the gas starts to flow along the surfaces of the target
disk.  Later, at $7.5 <t/t_0 < 15$, it flows strongly along the
 surfaces (see top panel of Figure 17).  The radial velocity of
matter flowing along the surfaces is $\sim(0.3-0.5) V_{ff}$.
Later, the layer became thicker owing to viscous dissipation and
the flow became more chaotic (Figure 17, bottom panel).  The
temperatures increased few times in the surface layer.  Here, we
should remind, that viscous heating is included, but cooling is
not.  If  radiative cooling balances viscous heating, then this
surface layer may survive much longer than  in the simulation.
The  surface flow is similar in some respects to that observed
in Model I (\S 3).  As in Model I, we also observed the increase
of the radial inflow velocity with decreasing $r$.

In this model, we also observed an interaction similar to that
investigated in Model II.  Namely, the interaction occurred not
only along the surfaces of the disk but also through the cross
section of the disk.  As a result, accretion was enhanced inside
the main disk as in Model II.  Compared with Model II, the
density of counterrotating matter is much smaller than that of
the ``target'' disk.  For this reason the interaction did not
lead to dramatic wave formation.  The radial velocity of viscous
flow inside the disk increased by a factor of $\sim 50$, and
matter flux increased by a similar factor.  However, the
velocity of accretion of the main disk remained much smaller
than the free-fall velocity.

Note,  that during this violent evolution of the outer layers of
the disk, the main part of the ``target'' disk was very stable,
the density did not change appreciably, and the radial
velocities remained small.

Figure 18 shows the time evolution of the matter flux through
the cylindrical surface $r=5 R_0$.  It  increases by a factor
$\sim 50$ compared with the initial  flux.  For $t/t_0
\lesssim10$ it increases mainly as a result of surface layer
accretion.  Later, there is also enhanced accretion from the
main part of the ``target'' disk.  A fraction about $\sim 0.2$
of the incoming matter is accreted.

\section{Two Applications}

Here, we briefly consider two applications of our results to
accretion disks.  One is related to galaxies and the second to
X-ray pulsars.

\subsection{Galaxies}

In some disk galaxies, counterrotating gas and/or stars are
observed (e.g., Rubin 1994a, 1994b; and Galletta 1995).  Even
though the present work treats Keplerian disks, we can
nevertheless make estimates based on our hydrodynamic results.
The rotation curves of galactic disks are flat ($v_\phi =$const)
over a significant range of distances owing in part to a
spheroidal distribution ($\rho \propto r^{-2}$) of dark matter.
Although this rotation curve is quite different from Keplerian,
the analytic viscous accretion flows are nearly the same for
flat and Keplerian rotation curves (LC).  The reason is that
the boundary layer formed between the oppositely rotating
components, where angular momentum is annihilated and fast
accretion occurs, does not depend strongly on the radial
dependence of $v_\phi$.

We consider the situation where a counterrotating cloud of
hydrogen of mass $M_{cl} = 10^9 M_\odot$ is captured into an
equatorial orbit of radius $r_{out}=20$kpc.  Model III is
pertinent to this situation.  Interaction between the
counterrotating cloud and existing corotating gas (assumed mass
$\geq M_{cl}$) in the galaxy leads to fast boundary layer
accretion along the top and bottom surfaces of the galactic disk
with average accretion speed $u_{CR} \sim 3.25 (h/H) V_c$, where
we have adapted the results of \S 3 to Model III which has two
boundary layers, and where we have used the circular velocity of
the galaxy $V_c\approx$ const rather than the Keplerian velocity
$V_K$.  The time-scale for this fast accretion to the center of
the galaxy is therefore

$$
t_{CR} \sim {r_{out}\over u_{CR}}\sim
3\times 10^8{\rm yr}
\left({r_{out} \over 20 {\rm kpc}}\right)
\left ({200 {\rm km/s} \over V_c }\right)\,,
\eqno(9{\rm a})
$$
where we have assumed $h/H ={\rm const}= 0.1$.  For the
reference values of equation (9a) and assuming that the cloud
mass is accreted in a time $\sim 10 t_{CR}$, the mass accretion
rate is

$$
\dot M \sim {2M_{cl} \over 10t_{CR}}
\sim 0.6~{\rm M}_\odot/{\rm yr}\,,
\eqno(9{\rm b})
$$
where the factor of two accounts for the fact that half of the
accreted gas is from the existing disk.  Note that the accretion
time-scale is much less than the Hubble time.

\subsection{X-Ray Pulsars}

Nelson et al. (1997a,b) and Chakrabarty et al. (1997) have
proposed that the change from spin-up to spin-down of X-ray
pulsars in binary systems results from a reversal of the angular
momentum of the wind supplied accreting matter.  The  matter
inflowing to the pulsar is assumed to be supplied at a constant
rate $\dot M$ at a radius $r_{out}$ in the equatorial plane of
the binary system.  Initially, the pulsar is assumed to have a
conventional, corotating disk.  Then, due to some disturbance,
the sense of rotation of the supplied matter reverses so that it
is counterrotating.  Model III is again pertinent to this
situation.  The `new' counterrotating matter encounters the
`old' corotating disk.  Interaction between the `new' and `old'
matter leads to fast boundary layer accretion with average
accretion speed $u_{CR} \sim 3.25(h/H) V_K$.  The time-scale for
this fast accretion to reach the pulsar is

$$
t_{CR}=  \int_0^{r_{out}} {dr \over u_{CR}}
\sim
2 {\rm d}
\left({ M_\odot \over M}\right)^{1\over 2}
\left({r_{out} \over 10^{12}{\rm cm}}
\right)^{3\over 2}\,,
\eqno(10)
$$
where we have again assumed
$h/H ={\rm const}= 0.1$.
The sign of the angular momentum carried by the boundary layer
flows at the much smaller Alfv\'en radius $r_A \ll r_{out}$,
which determines the spin-up or spin-down of the pulsar, is
uncertain from the present work.  If this has the sense of the
counterrotating material supplied at $r_{out}$, then a rapid
switch between spin-up and spin-down of the pulsar would occur
with time scale of order $t_{CR} \sim {\rm days}$.

On a much longer time-scale, inflowing counterrotating matter
will `use up' the old corotating disk.  This time scale is
simply the accretion time for a disk rotating in one direction,

$$
t_{SS}=\int_0^{r_{out}} { dr \over u_{SS}}
$$
$$
\sim 0.7 \times 10^5 {\rm d}
\left({0.1 \over \alpha}\right)
\left({0.01 \over c_s/V_K}\right)^2
\left({M_\odot \over M}\right)^{1\over 2}
\left({r_{out}\over 10^{12}{\rm cm}}\right)^{3\over 2}\,.
\eqno(11)
$$
A further time interval $\sim t_{SS}$ is required for
establishment of an equilibrium $\alpha-$disk rotating in the
opposite direction to the original disk.

\section{Conclusions}

Time-dependent, axisymmetric hydrodynamic simulations have been
used to study Keplerian accretion disks consisting of
counterrotating components.  The simulation code used for the
study has a numerical viscosity which is calibrated by study of
the accretion of a disk rotating in one direction.  Different
grid resolutions were used to study the influence of the
numerical viscosity.  Our simulation model does not include
radiative cooling.  Therefore, different values of the specific
heat ratio $\gamma$ from $1.01$ to $5/3$ were used in order to
assess the influence of heating due to the numerical viscosity.
A small value of $\gamma-1$ corresponds to almost isothermal
conditions.

Different cases were considered.  In Model I, the gas well above
the disk midplane rotates in one direction and that well below
has the same properties but rotates in the opposite direction.
In this case there is angular momentum annihilation in a narrow
equatorial boundary layer in which matter accretes
supersonically with a velocity which approaches the free-fall
velocity.  The average accretion speed of the disk can be
enormously larger than that for a conventional
$\alpha-$viscosity disk rotating in one direction.  For a much
lower viscosity (when we took a small part of the region and
calculated it with a grid resolution of $200\times200$), the
interface between the corotating and counterrotating components
shows significant warping, which is probably a type of
Kelvin-Helmholtz instability.  We observed that a large
viscosity suppresses this instability.

In Model II, we considered the case where the inner part of the
disk corotates while the outer part counterrotates.  In this
case a new equilibrium inner disk forms with a low density gap
between inner and outer disks.  In Model III we investigated the
case where low-density counterrotating matter inflowing from
large radial distances encounters an existing corotating disk.
Friction between the inflowing matter and the existing disk is
found to lead to fast boundary layer accretion along the disk
surfaces, while interaction of the disk with counterrotating
matter at large radii, leads to enhanced accretion in the main
body of the disk.  We observed that the boundary layer accretion
is a temporary phenomenon, because the interaction of the dense
disk and low-density counterrotating gas leads to heating of
this gas.  However, the interaction at large radii is more
steady and leads to continuous enhanced accretion in the main
disk.

These models are pertinent to the formation of counterrotating
disks in galaxies, in Active Galactic Nuclei, and in X-ray
pulsars in binary systems.  For galaxies the high accretion
speed allows counterrotating gas to be transported into the
central regions of a galaxy in a time much less than the Hubble
time.

It is clear that inclusion of the important role of
Kelvin-Helmholtz (KH) instabilities in counterrotating disks
requires $3$D simulations and this will be the subject of a
future paper.  For example, note that in Model I if the flow is
treated as a vortex sheet without radial inflow, then a local
stability analysis  ${\bf k}^2 H^2 \geq 1$ with ${\bf k}$ the
$(r,\phi)$ wavevector) indicates unstable KH warping for
wavenumbers $|k_\phi/k_r| < \sqrt{2}(c_s/V_K) \ll 1$ and a
maximum growth rate of $|k_r|c_s/2$ (LC; Choudhury \& Lovelace
1986).  To account for this non-axisymmetric warping instability
of the disk, we clearly need $3D$ simulations.  In Model II, a
local stability analysis of the interface between the inner and
outer disks indicates unstable KH warping for $|k_\phi/k_z| <
\sqrt{2} c_s/V_K \ll 1$ and a maximum growth rate of
$|k_z|c_s/2$.  Also in this case 3D simulations are needed.

A number of further developments are planned.  Simulations
explicitly including an $\alpha-$viscosity are desirable for
detailed comparison with theory.  Explicit inclusion of
radiative cooling is equally important.  Also, 3D simulations
are needed to  investigate the Kelvin-Helmholtz instability.

\acknowledgements {One of the authors (MMR) thanks Prof. R.
Giovannelli and D. Dale for helpful discussions.  We thank an
anonymous referee  for many constructive comments on an earlier
version of this work.  This work was supported in part by NSF
grant AST-9320068.  OAK and VMC were supported in part by the
Russian Federal Program ``Astronomy'' (subdivision ``Numerical
Astrophisics'') and by INTAS grant 93-93-EXT.  Also, this work
was made possible in part by Grant No. RP1-173 of the U.S.
Civilian R\&D Foundation for the Independent States of the
Former Soviet Union.  The work of RVEL was also supported in
part by NASA grant NAG5 6311.}

\end{document}